\pdfoutput=1
\documentclass[11pt,twoside,a4paper,cmspaper,final,collab]{cms-tdr}
\def\svnVersion{375da43}\def\svnDate{2023/09/05}\def\cmsCernNoTag{CERN-EP-2023-021}\def\cmsCernDate{\today}\def\cmsMessage{Published in the Journal of High Energy Physics as \href{http://dx.doi.org/10.1007/JHEP08(2023)204}{\doi{10.1007/JHEP08(2023)204}.}}
\begin{document}\cmsNoteHeader{TOP-22-012}

\newcommand{\sqs}{\ensuremath{\sqrt{s}}\xspace}
\newcommand{\sqrts}[1][13.6]{\ensuremath{\sqs=#1\TeV}\xspace}
\newcommand{\pp}{\ensuremath{\Pp\Pp}\xspace}
\newcommand{\intlumi}{\smash{\ensuremath{1.21\fbinv}}\xspace}

\newcommand{\statsyst}{\ensuremath{\,\text{(stat+syst)}}\xspace}
\newcommand{\pb}{\unit{pb}}
\newcommand{\xsecpred}{\ensuremath{924\hspace{1pt}^{+32}_{-40}\pb}\xspace}
\newcommand{\xsecmeas}{\ensuremath{881\pm 23\statsyst\pm 20\lum\pb}\xspace}

\newcommand{\sigeta}{\ensuremath{\eta}\xspace}
\newcommand{\abseta}{\ensuremath{\abs{\sigeta}}\xspace}
\newcommand{\mll}{\ensuremath{m_{\Pell\Pell}}\xspace}
\newcommand{\Deta}{\ensuremath{\Delta\sigeta}\xspace}
\newcommand{\Dphi}{\ensuremath{\Delta\phi}\xspace}

\newcommand{\ee}{\ensuremath{\EE}\xspace}
\newcommand{\emu}{\ensuremath{\Pepm\PGmmp}\xspace}
\newcommand{\mumu}{\ensuremath{\MM}\xspace}
\newcommand{\ljets}{\ensuremath{\Pell\text{+jets}}\xspace}
\newcommand{\ejets}{\ensuremath{\Pe\text{+jets}}\xspace}
\newcommand{\mjets}{\ensuremath{\PGm\text{+jets}}\xspace}

\newcommand{\Wjets}{\ensuremath{\PW\text{+jets}}\xspace}
\newcommand{\Zjets}{\ensuremath{\PZ\text{+jets}}\xspace}
\newcommand{\tW}{\ensuremath{\PQt\PW}\xspace}
\newcommand{\tchannel}{\ensuremath{t\text{ channel}}\xspace}

\newcommand{\hdamp}{\ensuremath{h_{\text{damp}}}\xspace}
\newcommand{\mtop}{\ensuremath{m_{\PQt}}\xspace}
\newcommand{\mZ}{\ensuremath{m_{\PZ}}\xspace}

\newcommand{\sigtt}{\ensuremath{\sigma_{\ttbar}}\xspace}
\newcommand{\muR}{\ensuremath{\mu_{\mathrm{R}}}\xspace}
\newcommand{\muF}{\ensuremath{\mu_{\mathrm{F}}}\xspace}

\cmsNoteHeader{TOP-22-012}
\title{First measurement of the top quark pair production cross section in proton-proton collisions at \texorpdfstring{\sqrts}{sqrt(s)=13.6 TeV}}

\date{\today}

\abstract{The first measurement of the top quark pair (\ttbar) production cross section in proton-proton collisions at \sqrts is presented. Data recorded with the CMS detector at the CERN LHC in Summer 2022, corresponding to an integrated luminosity of \intlumi, are analyzed. Events are selected with one or two charged leptons (electrons or muons) and additional jets. A maximum likelihood fit is performed in event categories defined by the number and flavors of the leptons, the number of jets, and the number of jets identified as originating from \PQb quarks. An inclusive \ttbar production cross section of \xsecmeas is measured, in agreement with the standard model prediction of \xsecpred.}

\hypersetup{%
    pdfauthor={CMS Collaboration},%
    pdftitle={First measurement of the top quark pair production cross section in proton-proton collisions at sqrt(s)=13.6 TeV},%
    pdfsubject={CMS},%
    pdfkeywords={CMS, top quark, Hadron-Hadron Scattering, Top Physics},%
}

\maketitle

\section{Introduction}

After an extended period of scheduled maintenance and upgrades beginning in late 2018, the CERN LHC resumed data taking in July 2022, and now operates at the unprecedented proton-proton (\pp) center-of-mass energy of 13.6\TeV.
This paper presents the first measurement of the top quark pair (\ttbar) production cross section \sigtt at the new energy.
Data corresponding to an integrated luminosity of \intlumi collected by the CMS experiment in Summer 2022 are analyzed.
The measurement marks the beginning of the third multi-year data-taking period of the LHC, which will provide data for new precision tests of the standard model (SM) of particle physics, allow for the continued exploration of the Higgs sector and of rare processes, and has the potential to reveal physics beyond the SM.

At the LHC, the \ttbar production cross section has been measured from \pp collisions by the ATLAS, CMS, and LHCb Collaborations at $\sqs=5.02$, 7, 8, and 13\TeV~\cite{CMS:TOP-11-007, ATLAS:2014nxi, CMS:TOP-14-018, CMS:TOP-12-006, CMS:TOP-13-004, ATLAS:2017jkf, ATLAS:2017wvi, LHCb:2018usb, CMS:TOP-17-001, CMS:TOP-18-005, ATLAS:2020ccu, ATLAS:2020aln, CMS:TOP-20-001, CMS:TOP-20-004, CMS:TOP-18-014, ATLAS:2022jbj, ATLAS:2022xny, ATLAS:2023gsl}.
The measurements are in agreement with predictions at next-to-next-to-leading order (NNLO) in perturbative quantum chromodynamics (QCD), which include resummation of the next-to-next-to-leading-logarithmic (NNLL) soft-gluon terms using the \textsc{top++} v2.0 program~\cite{Cacciari:2011hy, Barnreuther:2012wtj, Czakon:2012zr, Czakon:2012pz, Czakon:2013goa, Catani:2019iny, Czakon:2011xx}.
At \sqrts, assuming a top quark mass of $\mtop=172.5\GeV$, and using the PDF4LHC21 parton distribution function (PDF) set~\cite{PDF4LHCWorkingGroup:2022cjn}, a value of $\sigtt=\xsecpred$ is predicted, about 10\% larger than the value at \sqrts[13].

The measurement presented here provides a first test of whether \sigtt increases as expected at the new center-of-mass energy.
Novel methods are used to cross-check the initial calibration of the detector response to leptons and jets identified as originating from \PQb quarks (\PQb jets) using in situ constraints in a single fit to the \ttbar data sample.
The measurement also provides valuable information on the quality of the early 2022 data recorded by the CMS experiment, and allows for comparison to the Monte Carlo (MC) simulation.
Although the overall configuration of the CMS detector is largely unchanged since 2018, numerous upgrades have been installed, necessitating careful validation of the new data.

Events are selected with two charged leptons (electrons or muons, referred to as \Pell) of opposite charge (dilepton channel) or with a single lepton (\ljets channel).
In both channels, the presence of hadronic jets is required, and the \PQb jet identification improves the separation between signal and background events.
Events are further separated into categories defined by the number and flavor of the leptons, and the number of jets and \PQb jets.
The cross section is then extracted from a profile maximum likelihood fit to the event yields in the various categories.
In the fit, the combination of multiple event categories helps to constrain \PQb jet and lepton identification efficiencies, together with the other systematic uncertainties.
Tabulated results are provided in the HEPData record for this analysis~\cite{hepdata}.

\section{The CMS detector and event reconstruction}

The central feature of the CMS apparatus is a superconducting solenoid of 6\unit{m} internal diameter, providing a magnetic field of 3.8\unit{T}.
Within the solenoid volume are a silicon pixel and strip tracker, a lead tungstate crystal electromagnetic calorimeter (ECAL), and a brass and scintillator hadron calorimeter (HCAL), each composed of a barrel and two endcap sections.
The hadron forward (HF) calorimeter, made of steel and quartz-fibers, extends the pseudorapidity (\sigeta) coverage provided by the barrel and endcap detectors.
Muons are measured in gas-ionization detectors embedded in the steel flux-return yoke outside the solenoid.
A more detailed description of the CMS detector, together with a definition of the coordinate system used and the relevant kinematic variables, can be found in Ref.~\cite{CMS:Detector-2008}.

Events of interest are selected using a two-tiered trigger system.
The first level, composed of custom hardware processors, uses information from the calorimeters and muon detectors to select events at a rate of around 100\unit{kHz} within a fixed latency of about 4\mus~\cite{CMS:TRG-17-001}.
The second level, known as the high-level trigger (HLT), consists of a farm of processors running a version of the full event reconstruction software optimized for fast processing, and reduces the event rate to around 1\unit{kHz} before data storage~\cite{CMS:TRG-12-001}.

Numerous upgrades and improvements to the subdetectors, readout electronics, trigger, data-acquisition, software, and offline computing systems were implemented before the start of the 2022 data taking.
Notable examples include new silicon photomultipliers and readout electronics for the HCAL that allow for a finer granularity and longitudinal segmentation~\cite{CMS:TDR-010}, the replacement of the innermost layer of the silicon pixel detector~\cite{CMS:DP-2022-047}, the new hybrid CPU/GPU farm for the HLT~\cite{CMS:DP-2023-004}, and rebuilt dedicated online luminosity monitors~\cite{CMS:NOTE-2022-007, Karunarathna:2022tyd, Wanczyk:2022avs}.

The global event reconstruction with the particle-flow (PF) algorithm~\cite{CMS:PRF-14-001} aims to reconstruct and identify each individual particle in an event, with an optimized combination of all subdetector information.
In this process, the identification of the particle type (photon, electron, muon, charged hadron, neutral hadron) plays an important role in the determination of the particle direction and energy.
Photons are identified as ECAL energy clusters not linked to the extrapolation of any charged particle trajectory to the ECAL.
Each electron is identified as a charged particle track associated with one or more ECAL energy clusters, some of which may arise from bremsstrahlung photons emitted along the electron path through the tracker material.
Muons are identified as tracks in the central tracker consistent with either a track or several hits in the muon system, and associated with calorimeter deposits compatible with the muon hypothesis.
Charged hadrons are identified as charged particle tracks neither identified as electrons, nor as muons.
Finally, neutral hadrons are identified as HCAL energy clusters not linked to any charged hadron trajectory, or as a combined ECAL and HCAL energy excess with respect to the expected charged hadron energy deposit.
The primary vertex (PV) is taken to be the vertex corresponding to the hardest scattering in the event, evaluated using tracking information alone, as described in Section 9.4.1 of Ref.~\cite{CMS:TDR-15-02}.

For each event, hadronic jets are clustered from these reconstructed particles using the infrared- and collinear-safe anti-\kt algorithm~\cite{EXT:AK-2008, EXT:FastJet-2012} with a distance parameter of 0.4.
The jet momentum is determined as the vectorial sum of all particle momenta in the jet, and is found from simulation to be, on average, within 5--10\% of the true momentum over the entire jet transverse momentum (\pt) range used in the analysis~\cite{CMS:JME-13-004}.
Additional \pp interactions within the same or nearby bunch crossings, known as pileup, can contribute additional tracks and calorimetric energy depositions to the jet momentum.
The pileup-per-particle identification algorithm (PUPPI)~\cite{CMS:JME-18-001, EXT:PUPPI-2014} is used to mitigate the effect of pileup at the reconstructed-particle level, making use of local shape information, event pileup properties, and tracking information.
A local shape variable is defined, which distinguishes between collinear and soft diffuse distributions of other particles surrounding the particle under consideration.
The former is attributed to particles originating from the hard scatter and the latter to particles originating from pileup interactions.
Charged particles identified as originating from pileup vertices are discarded.
For each neutral particle, the local shape variable is computed using the surrounding charged particles compatible with the PV within the tracker acceptance ($\abseta<2.5$), and using all particles in the region outside the tracker coverage.
The momenta of the neutral particles are then rescaled according to their probability to originate from the PV deduced from the local shape variable, eliminating the need for jet-based pileup corrections~\cite{CMS:JME-18-001}.
Corrections to the jet energy scale (JES) are derived from simulation to bring the measured response of jets to that of particle-level jets on average.
In situ measurements of the momentum balance in dijet, {\PGg}+jet, {\PZ}+jet, and multijet events from data collected in 2022 are used to account for any residual differences in the JES between data and simulation~\cite{CMS:JME-13-004, CMS:DP-2022-054}.
Additional selection criteria are applied to each jet to remove jets potentially dominated by anomalous contributions from various subdetector components or reconstruction failures~\cite{CMS:JME-16-003}.

\section{Data and simulated samples}

The data were recorded between 27 July and 03 August 2022 using a combination of single-lepton and dilepton triggers that identify leptons within $\abseta<2.5$.
The single-lepton HLT selection requires the presence of an isolated electron (muon) reconstructed with $\pt>32$ (24)\GeV.
The dilepton HLT selection requires the presence of two isolated electrons with $\pt>23$ and 12\GeV, two isolated muons with $\pt>17$ and 8\GeV, one isolated electron with $\pt>23\GeV$ and one isolated muon with $\pt>8\GeV$, or one isolated muon with $\pt>23\GeV$ and one isolated electron with $\pt>12\GeV$.
The trigger efficiencies in data and simulation are determined in intervals of \pt and \abseta using a ``tag-and-probe'' method~\cite{CMS:EWK-10-002} on a sample of events with leptonically decaying \PZ bosons.
The trigger efficiencies in simulation are corrected to match the data.

The CMS detector recorded \intlumi of data during this time period that pass data quality requirements, yielding the data set used in this measurement.
The integrated luminosity is measured using transverse energy deposits in the HF, which has been calibrated based on a preliminary analysis of the data recorded during the van der Meer scan program~\cite{vanderMeer:296752, Grafstrom:2015foa, CMS:LUM-17-003} performed in November 2022.
The average number of \pp interactions per bunch crossing in the data sample is around 40.

Simulated event samples, produced using MC event generators, are used to evaluate the signal selection efficiency and to predict the contributions from background processes.
Events for the \ttbar signal process, as well as for single top quark production in the $t$ and \tW channels, are simulated with the \POWHEG v2 generator~\cite{EXT:Powheg-ttbar-2007, EXT:Powheg-2007, EXT:Powheg-2010} at next-to-leading order in QCD.
Top quark decays are simulated with \POWHEG for the \ttbar and \tW-channel single top quark samples, and with \textsc{MadSpin}~\cite{EXT:Madspin-2013} for the $t$-channel single top quark sample.
Events for $\PZ/\PGg^\ast$ and \PW boson production with up to four additional jets (\Zjets and \Wjets) are simulated with the \MGvATNLO v2.6.5 generator~\cite{EXT:MG5aMCatNLO-2014} at leading order in QCD.
Diboson production events ($\PW\PW$, $\PW\PZ$, and $\PZ\PZ$) are simulated with the \PYTHIA8.306 program~\cite{EXT:Pythia-2015} at leading order in QCD.

For all simulated samples, the proton structure in the matrix-element (ME) calculation is described with the NNPDF3.1 PDF set~\cite{EXT:NNPDF-2017} at NNLO.
The parton showering, hadronization, and underlying event are simulated using \PYTHIA8 with the CP5 tune~\cite{CMS:GEN-17-001}.
For the samples simulated with \POWHEG, the \hdamp parameter, which regulates the parton shower (PS) matching scale, is set to 1.379\mtop~\cite{CMS:GEN-17-001}.
For the samples simulated with \MGvATNLO, the PS matching is performed with the MLM prescription~\cite{EXT:Merging-2008}.
The CMS detector response is simulated with the \GEANTfour toolkit~\cite{EXT:GEANT4-2002}.
To model the effect of pileup, additional minimum bias interactions are superimposed on the simulated hard-scatter events, with a multiplicity matching that inferred from data.

All simulated samples are normalized to the product of the corresponding theoretical cross section and the integrated luminosity of the data sample.
Top quark production cross sections are calculated assuming a top quark mass of $\mtop=172.5\GeV$ and are evaluated using the PDF4LHC21 PDF set~\cite{PDF4LHCWorkingGroup:2022cjn}.
The \ttbar signal samples are normalized to the NNLO+NNLL cross section of \xsecpred.
Single top quark production in the \tchannel is normalized to the cross section calculated at NNLO in QCD with \MCFM v10.1~\cite{Campbell:2020fhf}, and in the \tW channel to the cross section calculated up to approximate next-to-NNLO corrections matched to complete NLO results in QCD, as described in Ref.~\cite{Kidonakis:2021vob}. 
For \Wjets and \Zjets production, cross sections have been calculated at NNLO in QCD with the \textsc{DYTurbo} v1.2 program~\cite{Camarda:2019zyx}, and for diboson production at next-to-leading order in QCD with \textsc{matrix} v2.1.0~\cite{Grazzini:2017mhc}.

\section{Event selection}

In the analysis, reconstructed electrons or muons are considered from the range $\abseta<2.4$.
Electrons reconstructed with $1.44<\abseta<1.57$, in the transition region between the barrel and endcap regions of the ECAL, are removed because of the suboptimal electron reconstruction in this region.
We use additional identification (ID) criteria to identify ``prompt'' leptons originating from top quark decays at the PV and to reduce background contributions from ``nonprompt'' leptons, defined as any reconstructed lepton that does not match a true lepton produced in a hard scattering process at the PV.
The latter category includes leptons from photon conversions, leptons from semileptonic decays of hadrons, and hadrons misidentified as leptons.

For electrons, we apply the ``tight'' working point of the cut-based ID criteria described in Ref.~\cite{CMS:EGM-17-001}.
These criteria include requirements on the shape of the electromagnetic shower in the ECAL, the matching between the track and the ECAL cluster, and the track quality.
The muon ID criteria correspond to the ``tight'' working point described in Ref.~\cite{CMS:MUO-16-001}.
These include requirements on the quality and the matching of the tracks reconstructed in the inner tracker and in the muon detectors, and on the compatibility with originating from the PV.

The relative isolation variable is defined as the \pt sum of all reconstructed PF candidates (except the lepton itself) within a cone of fixed radius around the lepton direction, divided by the lepton \pt~\cite{CMS:PRF-14-001}.
The cone radius is defined in terms of the separation variable $\DR=\sqrt{\smash[b]{(\Deta)^2+(\Dphi)^2}}$, where \Deta and \Dphi are the \sigeta and azimuthal angle differences.
For electrons, the ID criteria in Ref.~\cite{CMS:EGM-17-001} include a requirement on the relative isolation calculated with $\DR<0.3$.
For muons, we apply the ``tight'' requirement from Ref.~\cite{CMS:MUO-16-001} on the relative isolation calculated with $\DR<0.4$.
In both cases, corrections for the contributions from pileup particles to the isolation sum are applied~\cite{CMS:EGM-17-001, CMS:MUO-16-001}.

Jets with $\pt>30\GeV$ and $\abseta<2.4$ are selected for this analysis, and are required to be separated from any selected lepton by $\DR>0.4$.
The \textsc{DeepJet} algorithm~\cite{CMS:BTV-16-002, CMS:DP-2018-058, EXT:DeepJet-2020}, trained on simulated events at \sqrts[13], is used for the identification of \PQb jets (``\PQb tagging'').
We apply a working point with an observed selection efficiency for \PQb quark jets of more than 80\%, and a misidentification rate for \PQc quark jets (light quark and gluon jets) of around 17 (2)\%, as determined in simulated \ttbar events.

For the \ljets channel, events are selected by requiring exactly one lepton with $\pt>35\GeV$ in order to match the single-electron trigger threshold.
Events with additional leptons with $\pt>10\GeV$ are removed.
At least three jets are required, with either one or two jets passing the \PQb tagging requirements.
The selected events in the \ljets channel are divided into the two flavor categories:\ \ejets and \mjets.

In the dilepton channel, events are selected by requiring exactly two leptons of opposite charge and at least one jet.
The highest \pt (``leading'') and second-highest \pt lepton are each required to have $\pt>35\GeV$.
Events with additional lepton candidates of $\pt>10\GeV$ are removed.
We further require that the invariant mass \mll of the two leptons is larger than 20\GeV to remove background contributions from low-mass resonances.
The selected events in the dilepton channel are divided into the three flavor categories:\ \emu, \ee, and \mumu.

In the \ee and \mumu channels, the reconstructed \mll is required to differ from the \PZ boson mass $\mZ=91.2\GeV$~\cite{EXT:PDG-2022} by more than 15\GeV, in order to suppress background contributions from \Zjets production.
In these channels it is also required that at least one of the selected jets is identified as a \PQb jet.

The five flavor categories are further divided by the number of selected \PQb jets.
For the \ee, \mumu, \ejets, and \mjets events, two categories each are formed with exactly one or exactly two \PQb jets.
For the \emu channel, three categories are formed with zero, exactly one, or exactly two \PQb jets.
A total of 11 event categories are obtained, which are further subdivided by the total number of jets yielding 40 bins to be considered in the maximum likelihood fit.

\section{Background estimation}

\begin{figure}[!p]
\centering
\includegraphics[width=0.49\textwidth]{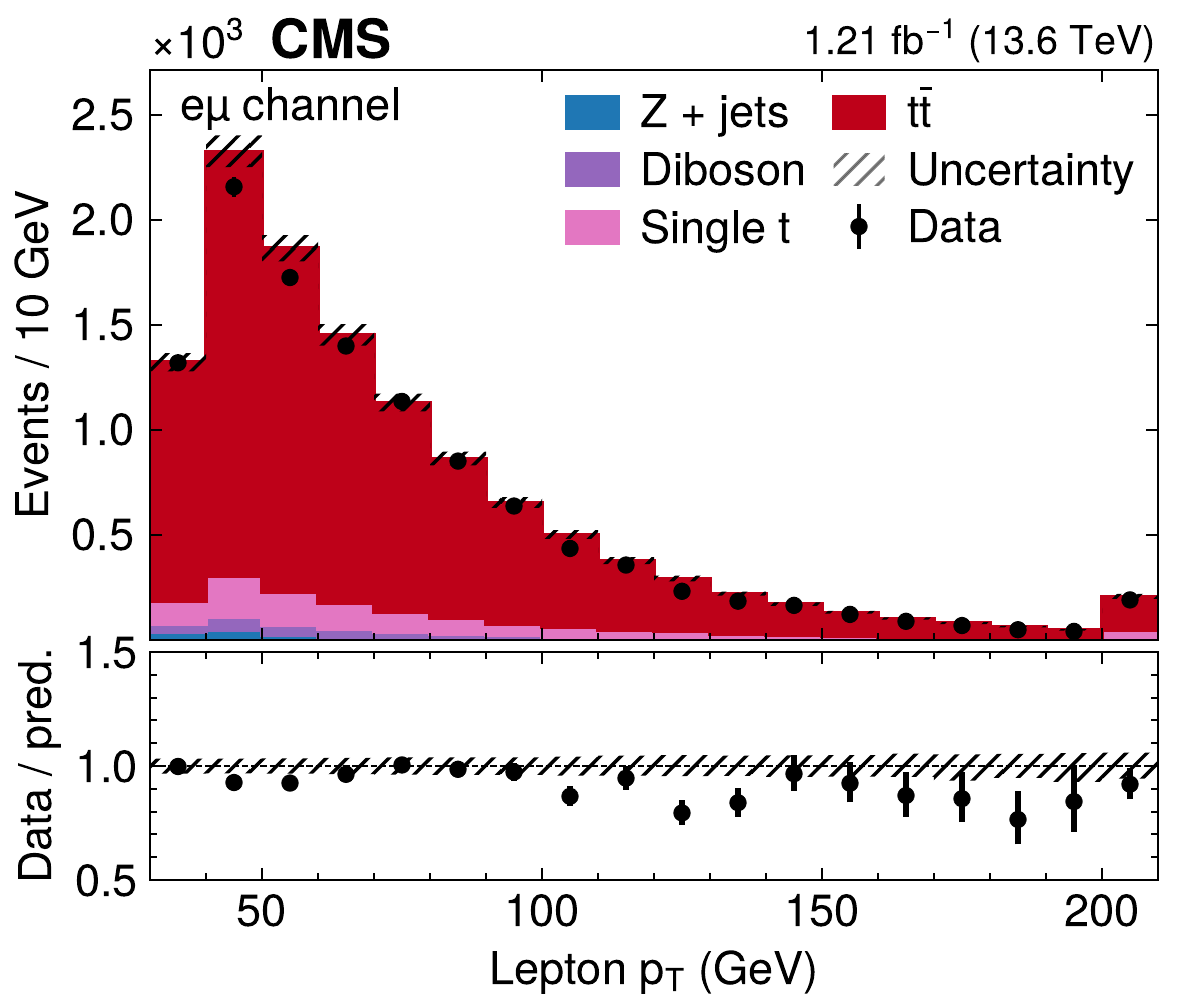}
\hfill
\includegraphics[width=0.49\textwidth]{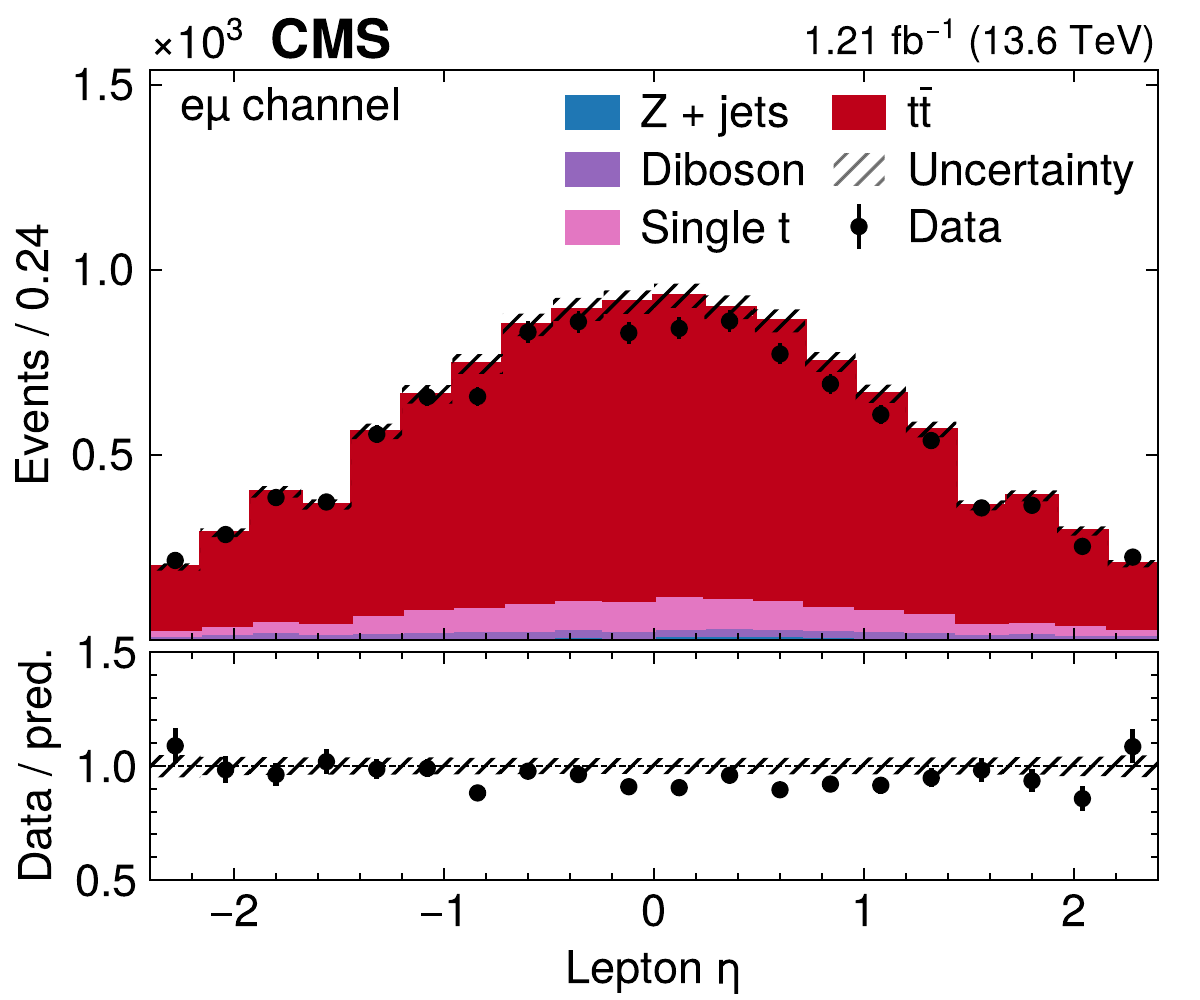}
\includegraphics[width=0.49\textwidth]{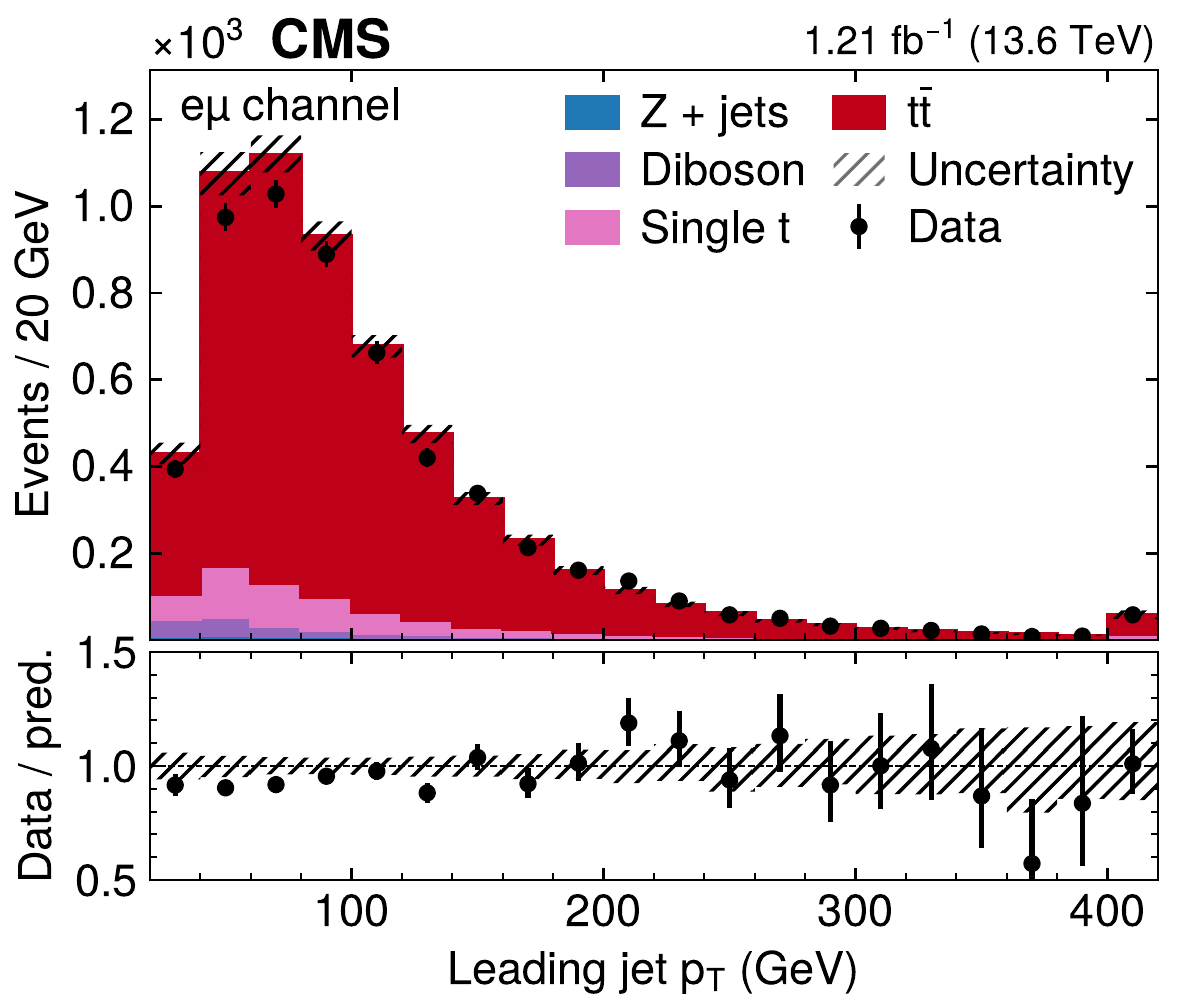}
\hfill
\includegraphics[width=0.49\textwidth]{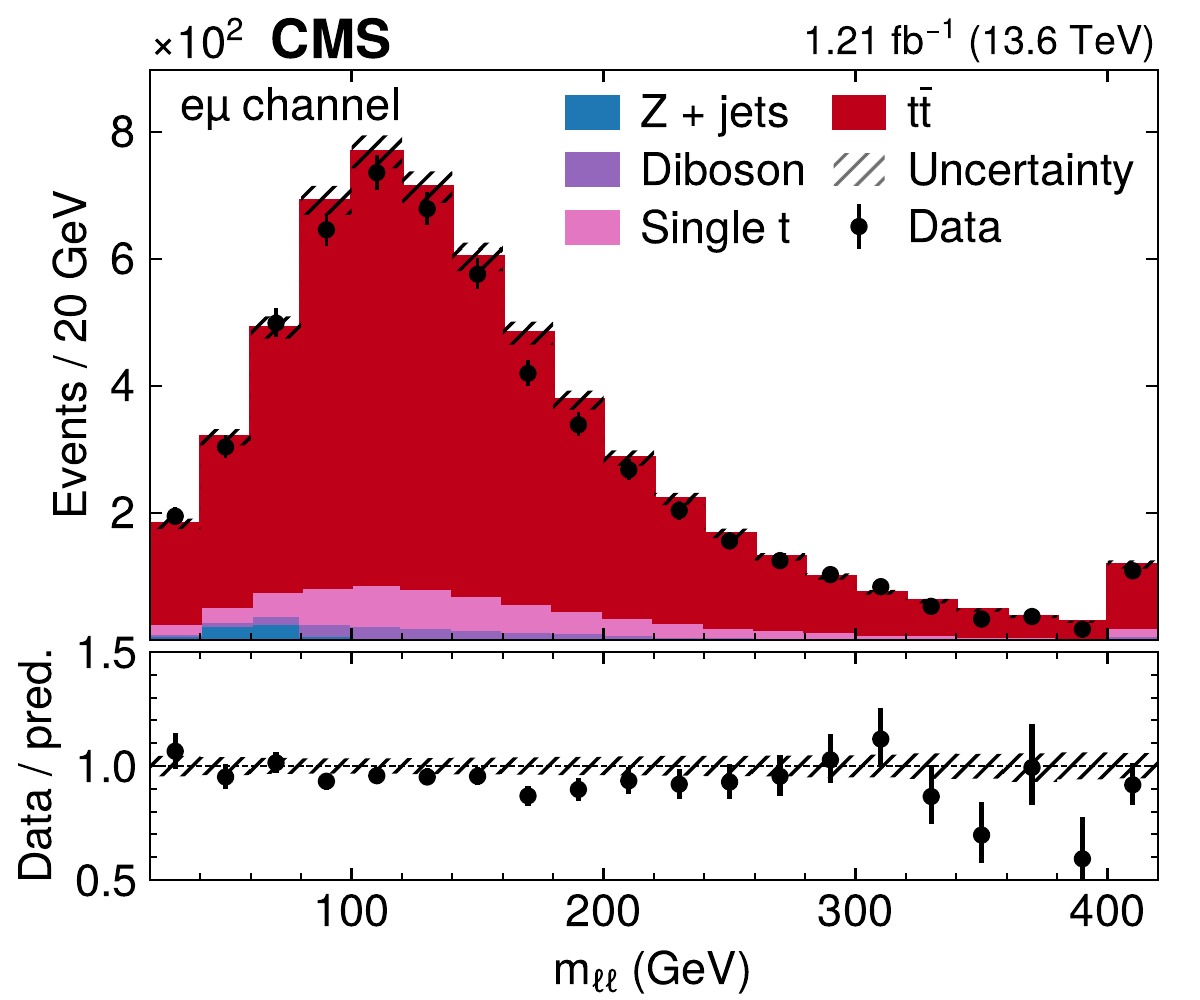}
\includegraphics[width=0.49\textwidth]{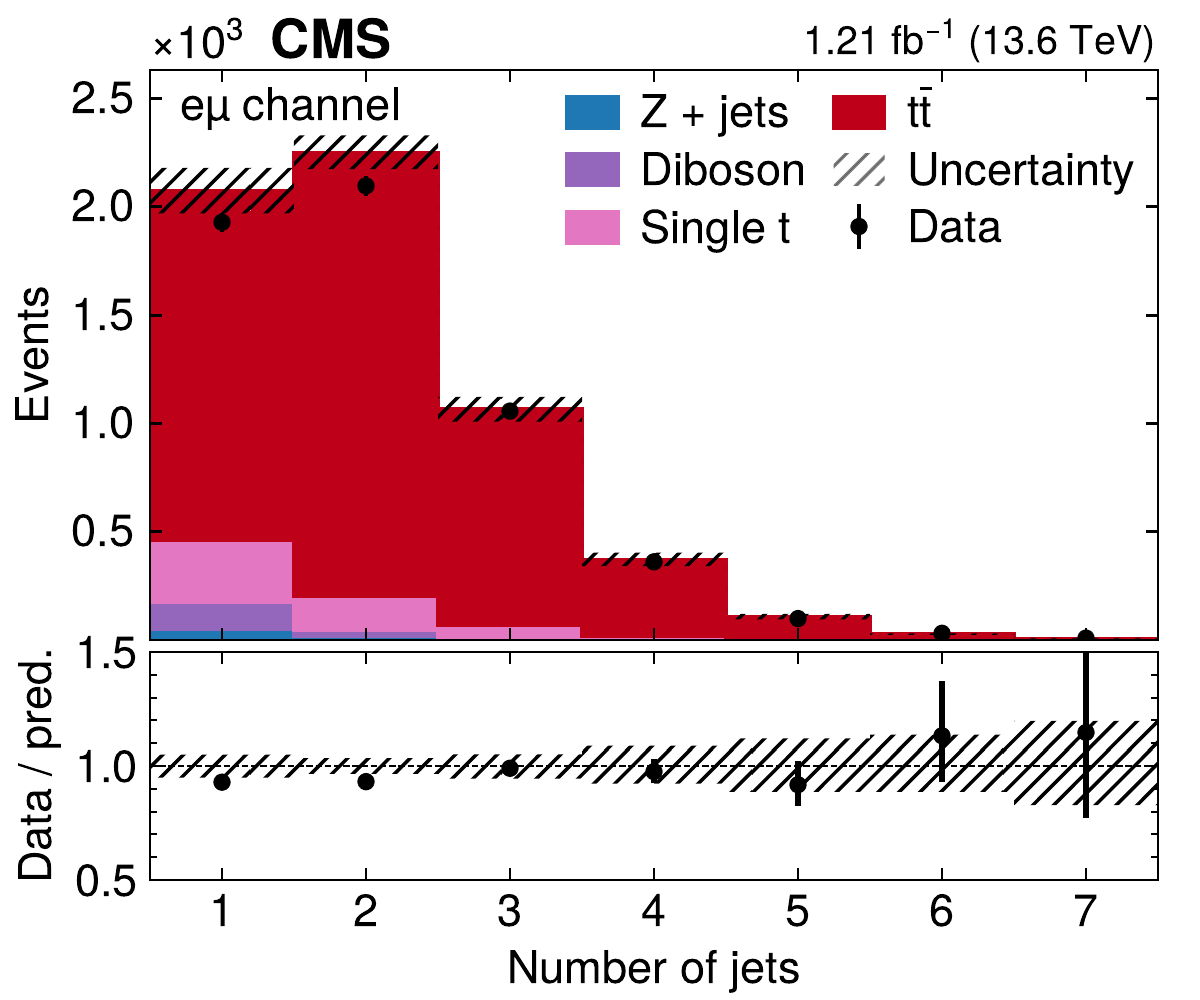}
\hfill
\includegraphics[width=0.49\textwidth]{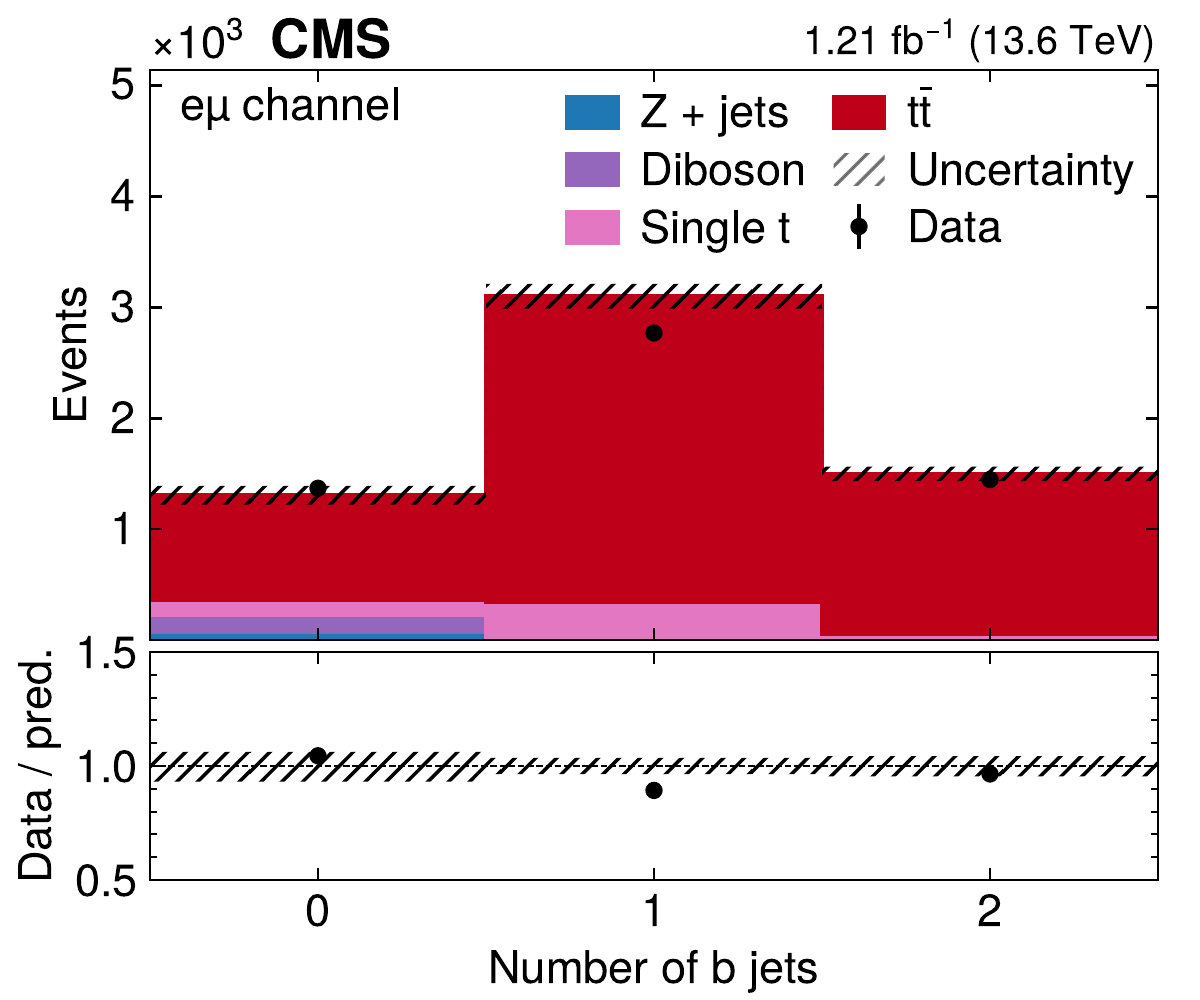}
\caption{%
    Comparison of the number of observed (points) and predicted (filled histograms) events in the \emu channel.
    The distributions of the \pt (upper left) and \sigeta (upper right) of both leptons, the leading jet \pt (middle left), \mll (middle right), and the number of jets (lower left) and \PQb jets (lower right) are displayed.
    The predictions are normalized using the measured integrated luminosity and predicted cross sections, and are scaled by the \PQb jet scale factors as obtained from the fit.
    The vertical bars on the points represent the statistical uncertainties in the data, and the hatched bands the systematic uncertainty in the predictions, including the integrated luminosity.
    The last bins include the overflow contributions.
    In the lower panels, the ratio of the event yields in data to the sum of predicted signal and background yields is presented.
}
\label{fig:em}
\end{figure}

\begin{figure}[!p]
\centering
\includegraphics[width=0.49\textwidth]{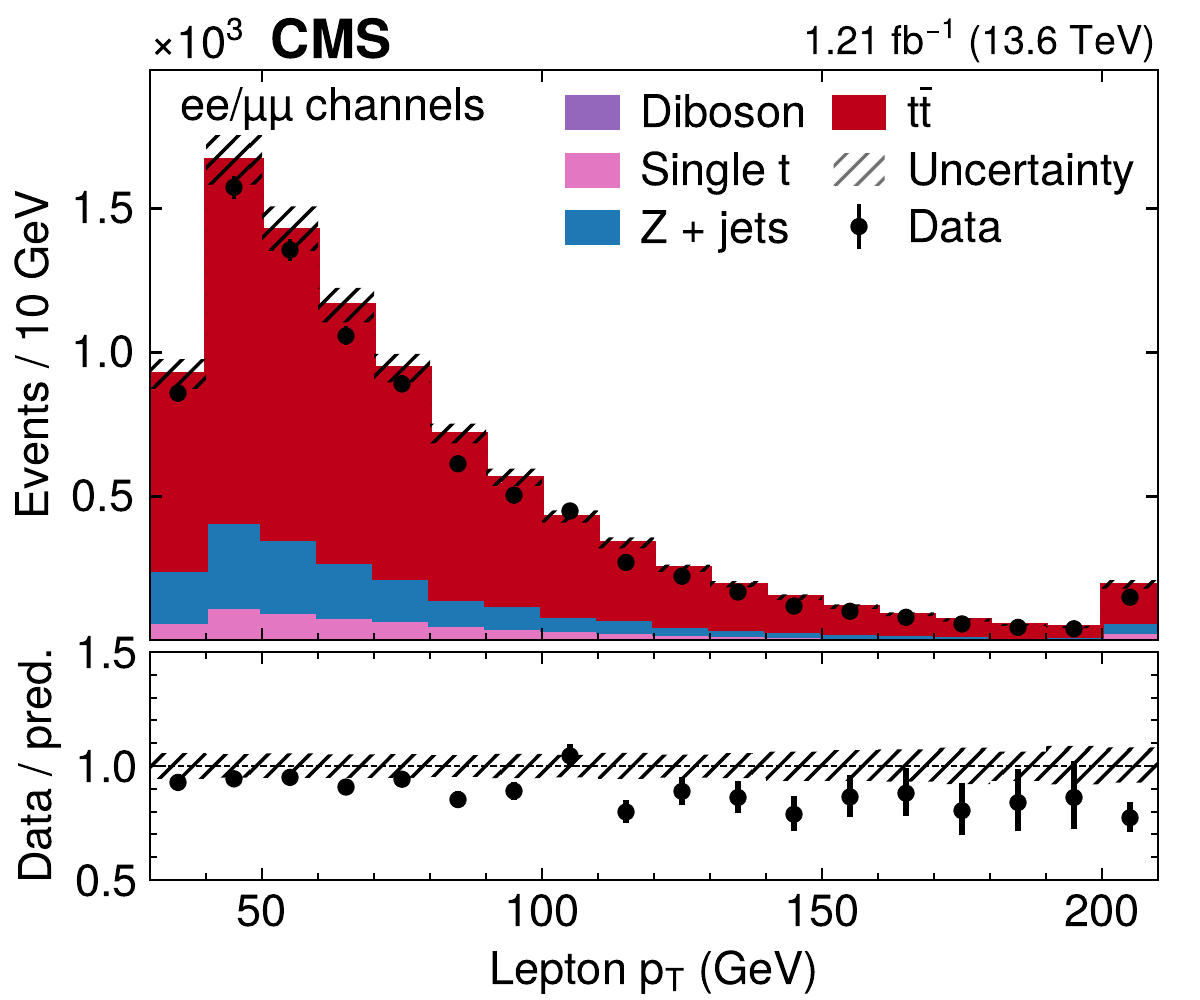}
\hfill
\includegraphics[width=0.49\textwidth]{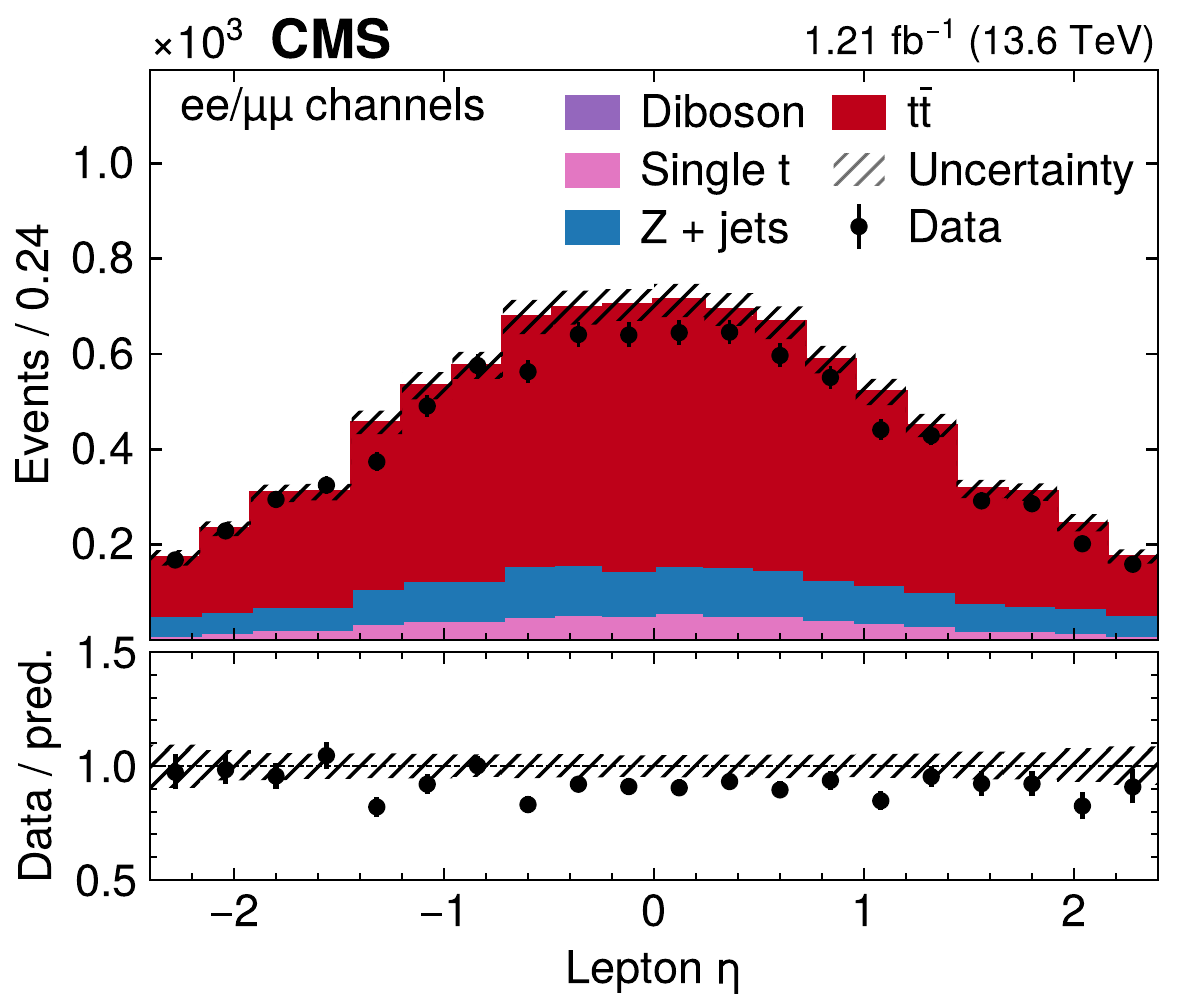}
\includegraphics[width=0.49\textwidth]{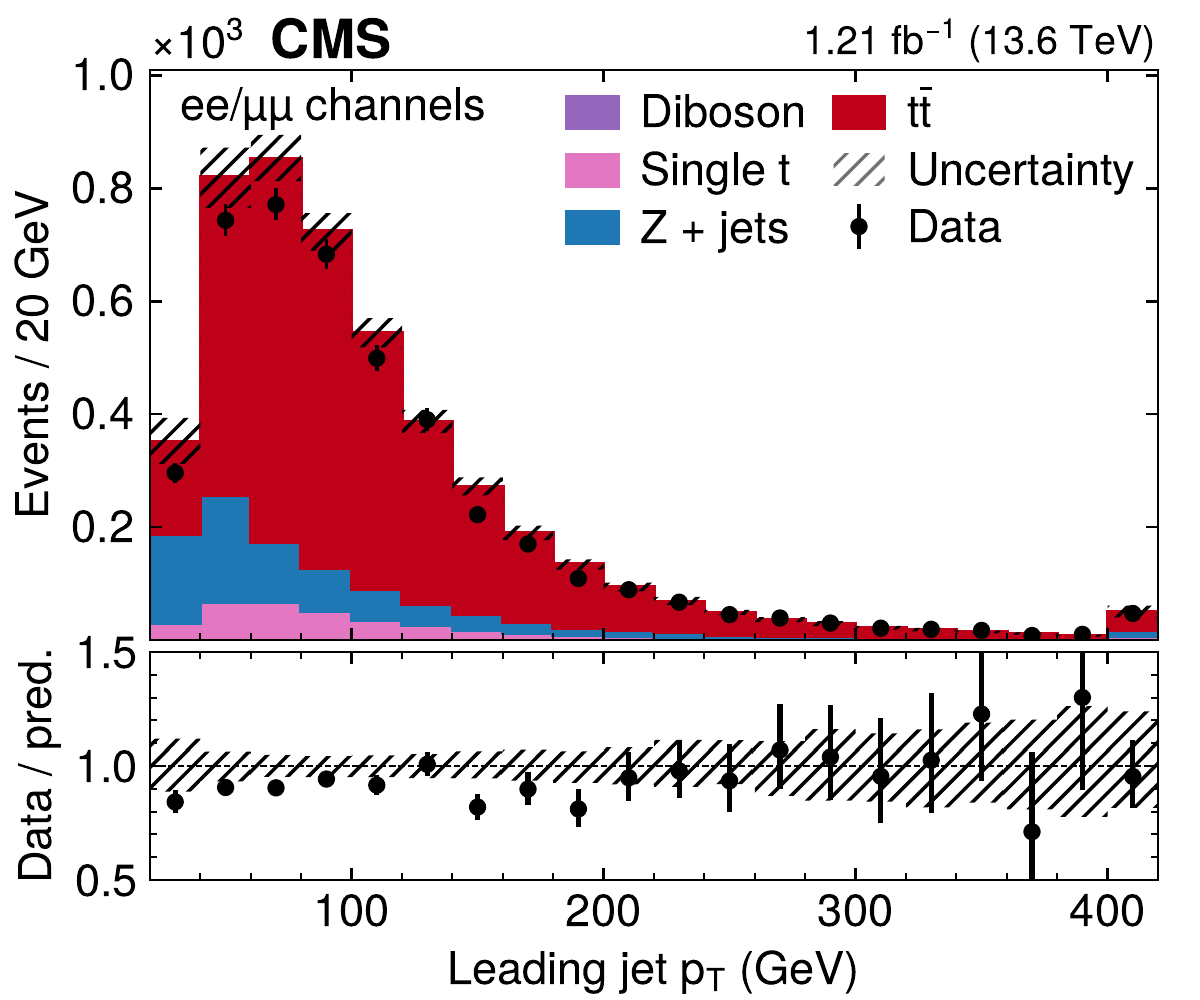}
\hfill
\includegraphics[width=0.49\textwidth]{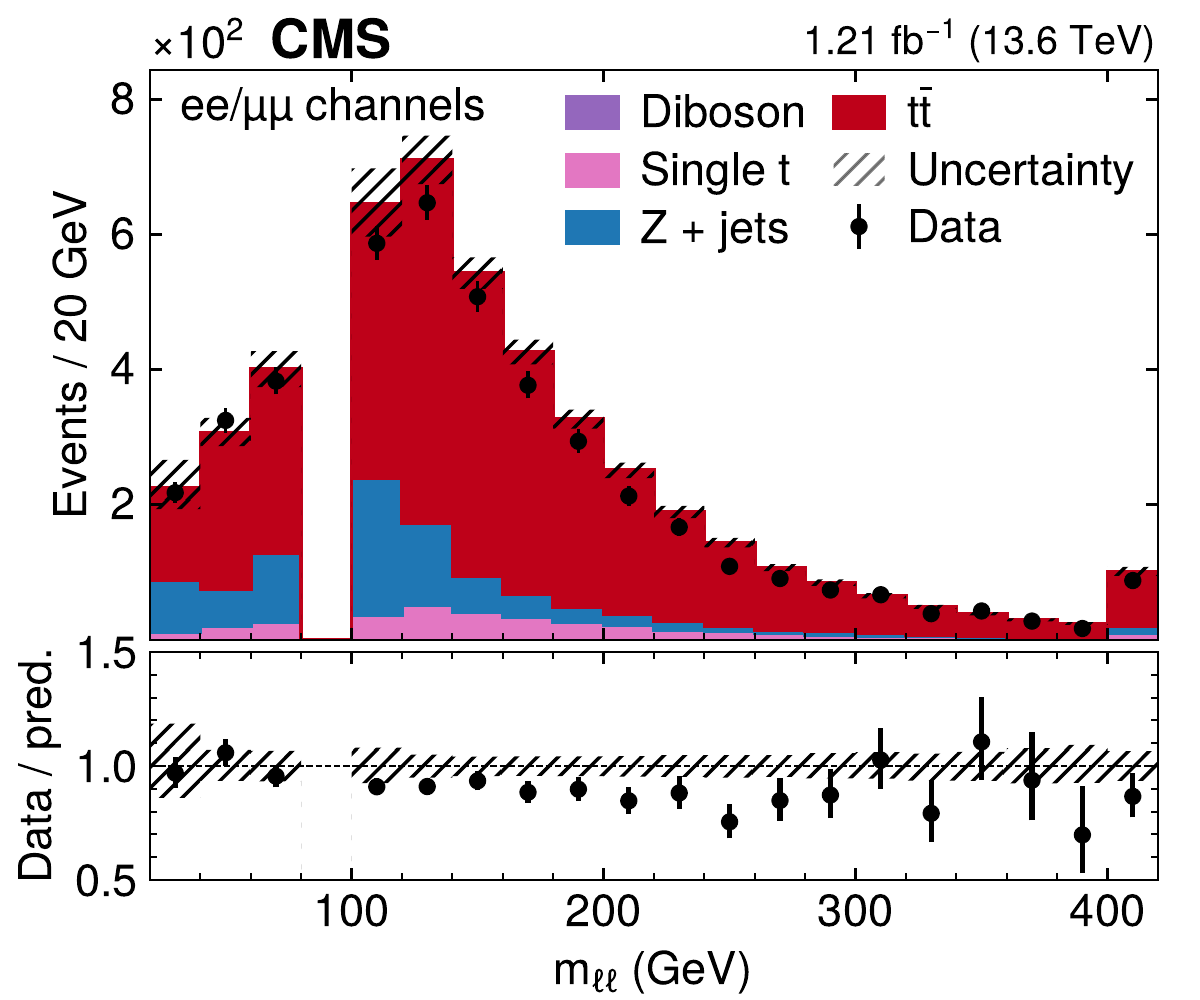}
\includegraphics[width=0.49\textwidth]{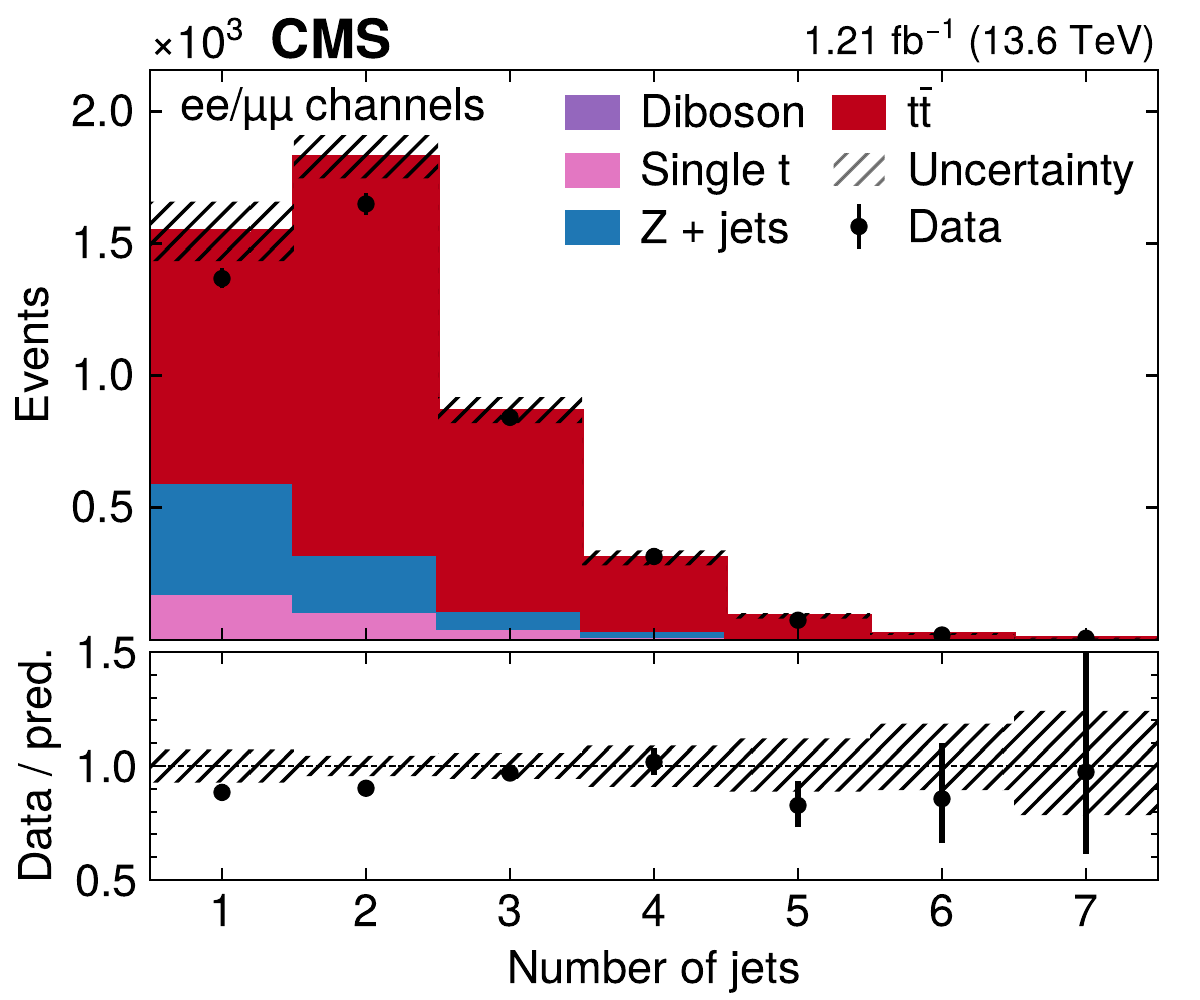}
\hfill
\includegraphics[width=0.49\textwidth]{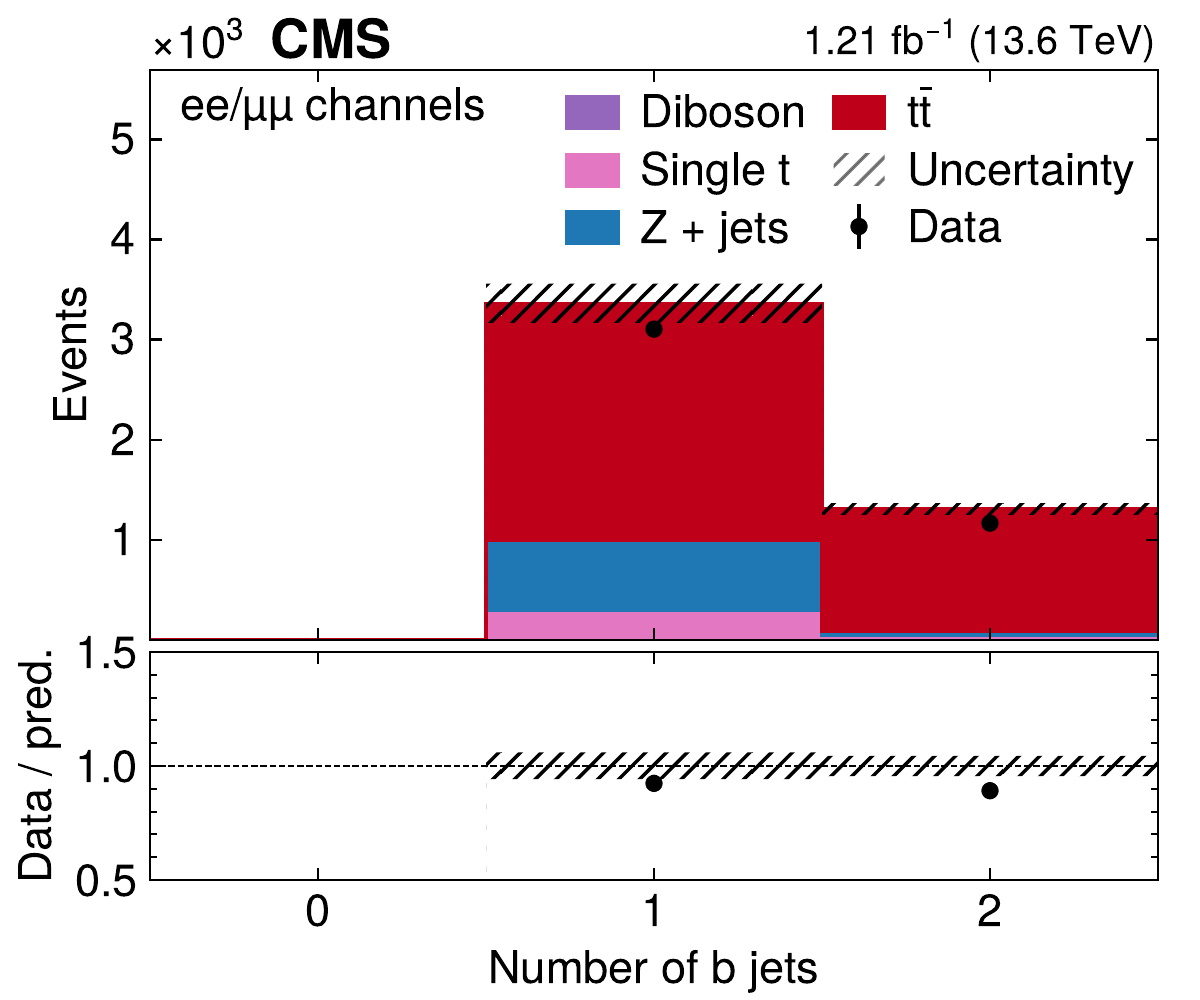}
\caption{%
    The number of observed and predicted events in the \ee and \mumu channel are presented in the same manner as Fig.~\ref{fig:em}.
}
\label{fig:eemm}
\end{figure}

\begin{figure}[!p]
\centering
\includegraphics[width=0.49\textwidth]{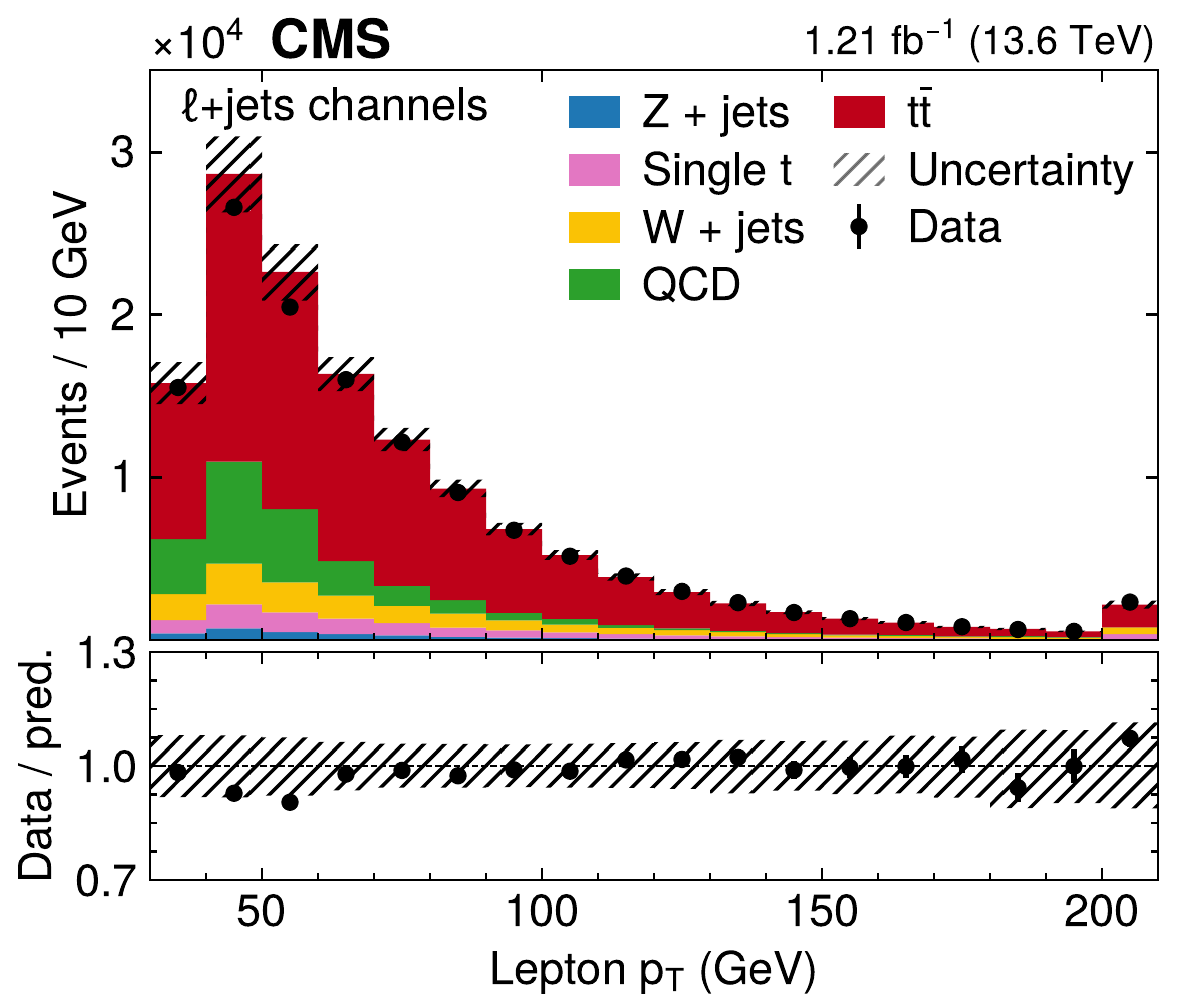}
\hfill
\includegraphics[width=0.49\textwidth]{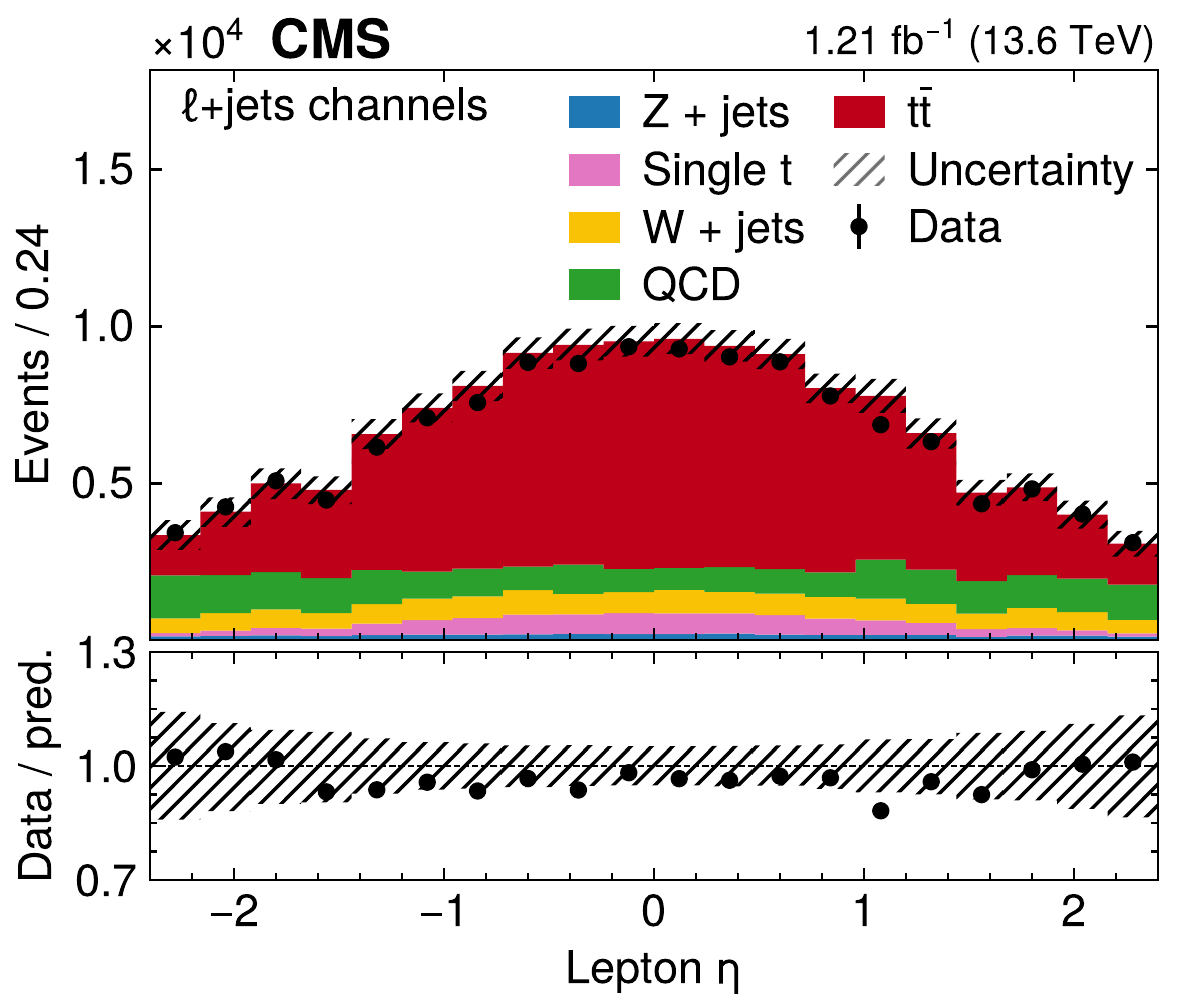}
\includegraphics[width=0.49\textwidth]{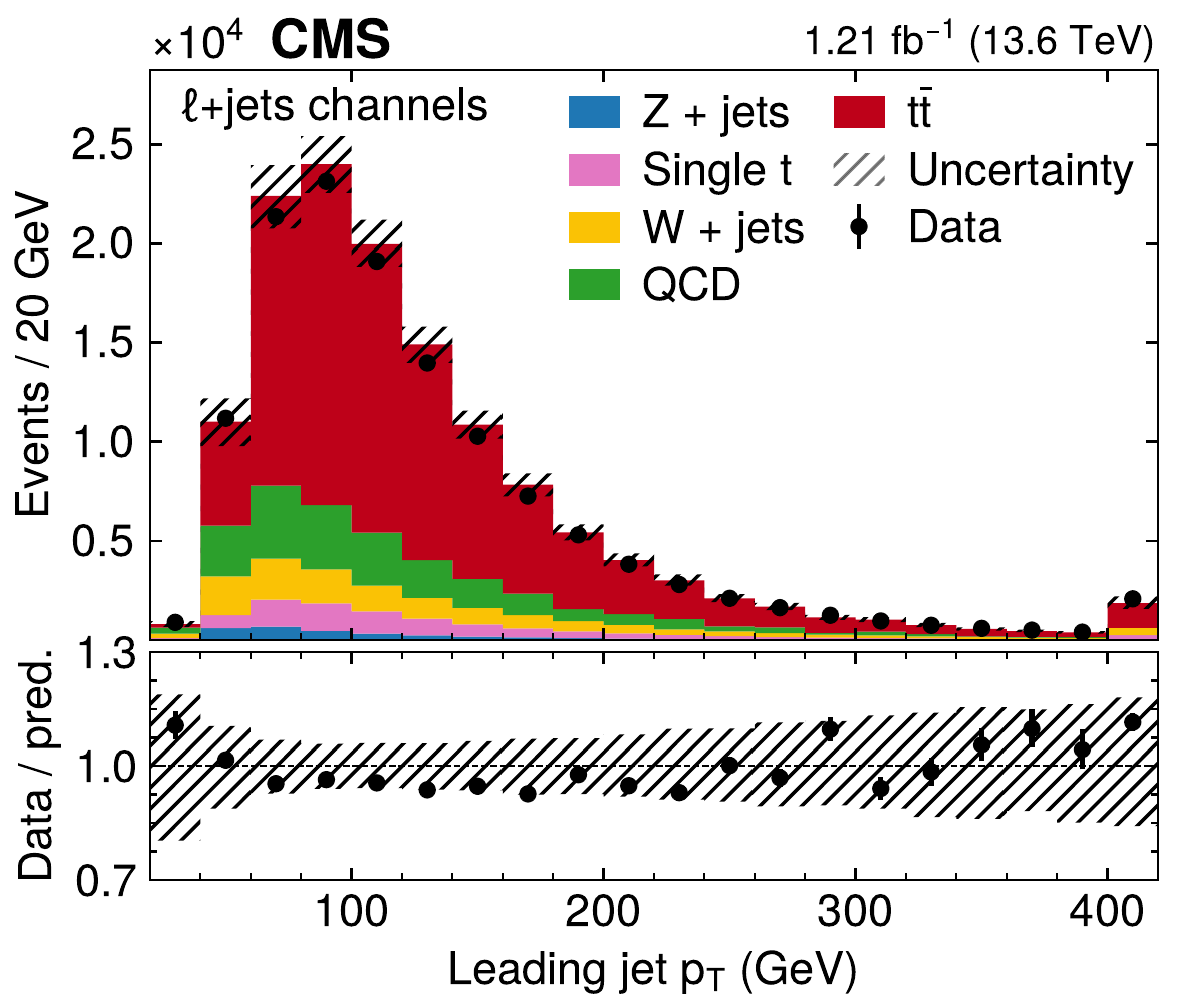}
\hfill
\includegraphics[width=0.49\textwidth]{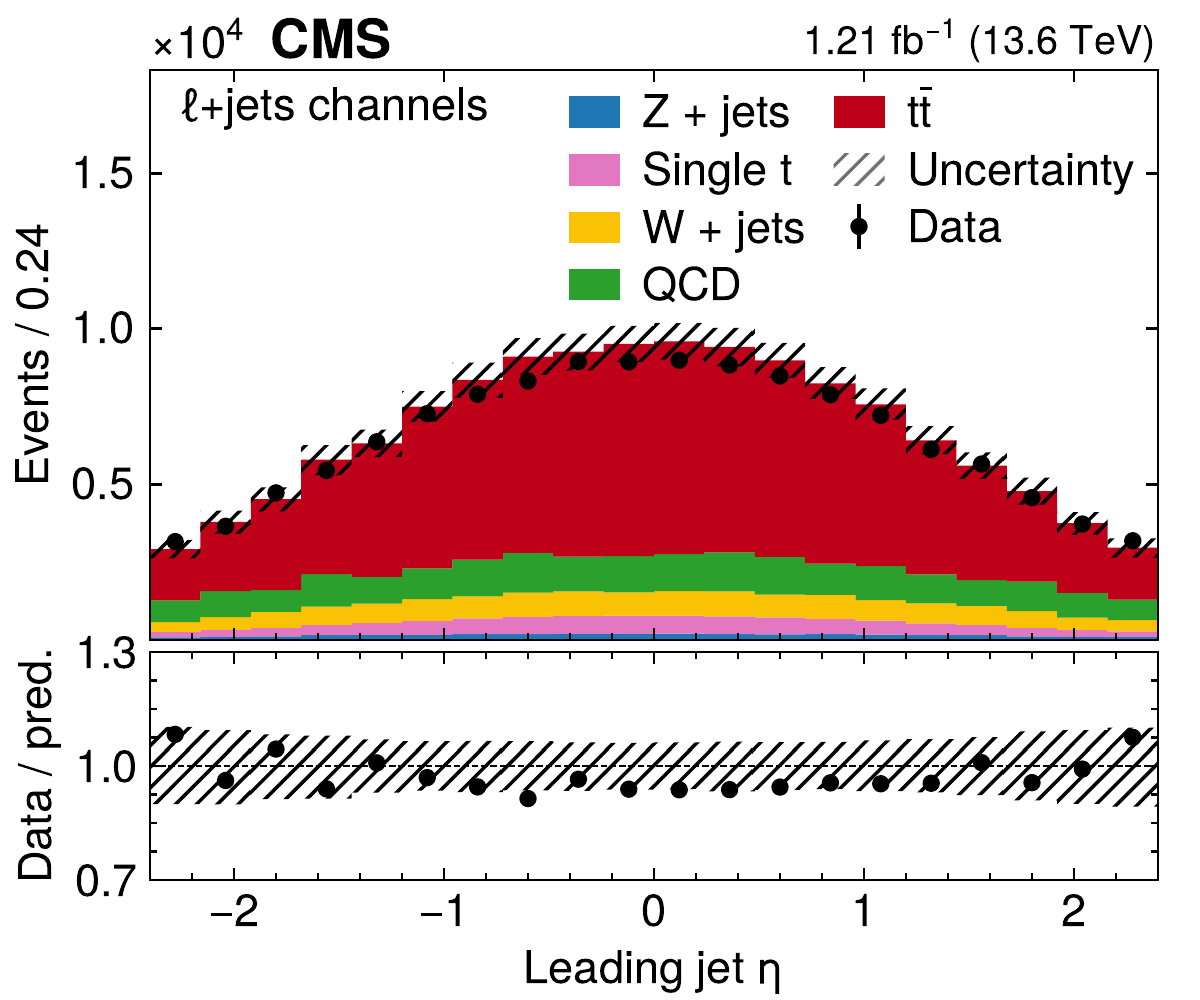}
\includegraphics[width=0.49\textwidth]{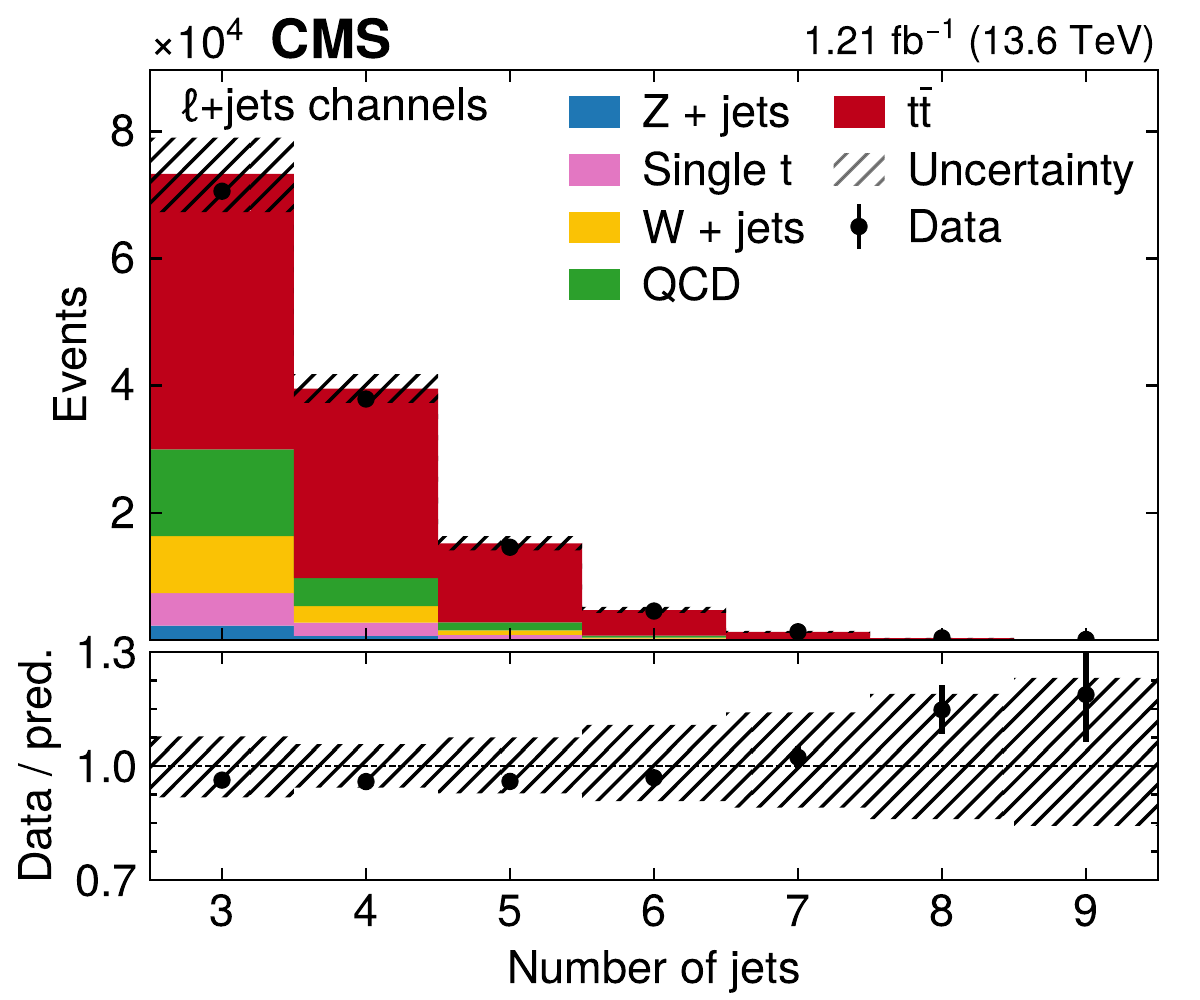}
\hfill
\includegraphics[width=0.49\textwidth]{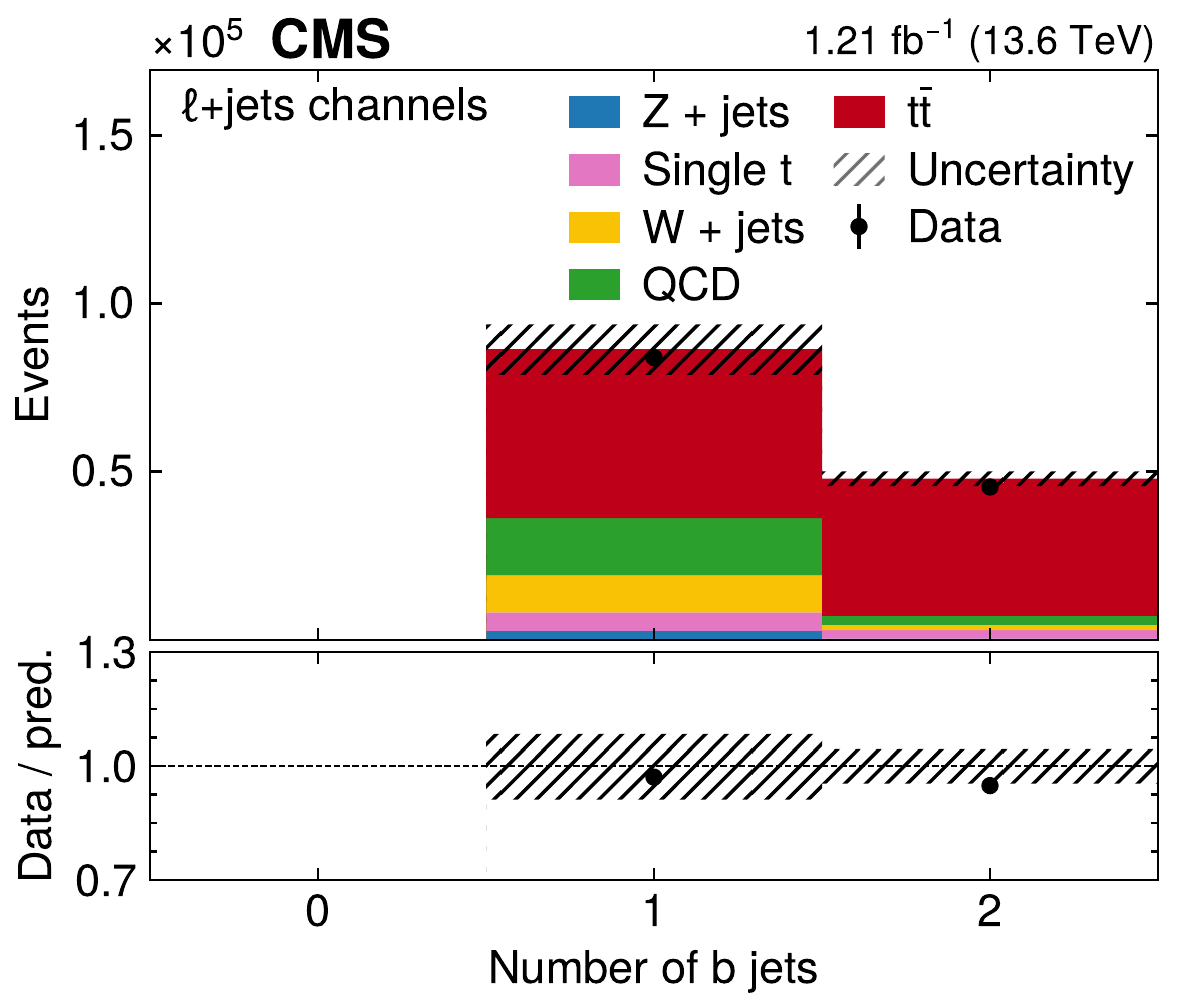}
\caption{%
    The number of observed and predicted events in the \ljets channel are presented in the same manner as Fig.~\ref{fig:em}, except that the middle-right plot shows the \sigeta of the leading jet instead.
}
\label{fig:ljets}
\end{figure}

Background contributions after the event selection arise from single top quark (\tchannel and \tW, collectively referred to as ``single \PQt''), \Zjets, \Wjets, and diboson production.
They are estimated using simulated event samples.
Additional contributions arise from SM events composed uniquely of jets produced through the strong interaction, referred to as QCD multijet events, which are estimated using control regions (CRs) in data.
The resulting agreement between data and simulation, prior to performing the fit, is shown in Figs.~\ref{fig:em}--\ref{fig:ljets}.
Efficiency corrections related to \PQb jet identification, as determined in the fit and described in Sections~\ref{sec:systematics}--\ref{sec:fit}, are applied to the simulation in these figures.
The largest background contributions are single-\PQt production in the \emu channel, \Zjets production in the \ee and \mumu channels, and QCD multijet events in the \ljets channels.

The predicted background from \Zjets production depends strongly on the number of \PQb jets passing the selection, but the distribution of additional \PQb jets is not always well-modeled in simulation.
To account for this, a method based on control samples in data is used to determine the best normalization for the \Zjets sample after event selection.
A CR consisting of dilepton events with $\abs{\mll-\mZ}<15\GeV$ is studied following the methods of Ref.~\cite{CMS:EXO-16-049}.

QCD multijet events contribute to the \ljets final state because of the presence of nonprompt leptons, whereas their contribution is negligible in the dilepton channels.
We estimate the QCD multijet background in the \ljets channel from two orthogonal CRs in data.
First, we select events with leptons that fail the relative isolation requirement while still passing all other ID requirements, and use this region to derive a shape template for the QCD contribution.
We then define a second CR consisting of \ljets events with only one selected jet, which must be identified as a \PQb jet.
From this CR, we determine the normalization of the shape template (coarsely binned in lepton \pt and \abseta) by measuring the ratio of event counts passing or failing the relative lepton isolation requirement.
The resulting estimation of QCD multijet contributions leads to good agreement with the data, as shown in Fig.~\ref{fig:ljets}.

\section{Systematic uncertainties}
\label{sec:systematics}

Nuisance parameters are used to describe the systematic uncertainties in the fit procedure, described in Section~\ref{sec:fit}.
For each uncertainty source, the simulated event samples are used to construct corresponding template histograms that describe the expected signal and background distributions for a given nuisance parameter variation.
In the fit of the templates to the data, the best values for the parameter of interest (which is \sigtt) and all nuisance parameters are determined simultaneously.

Lepton ID efficiencies are derived as functions of lepton \abseta, \pt, and flavor using the tag-and-probe method in a \Zjets-enriched CR.
Scale factors are applied to simulated event samples to match the efficiencies in data, and the uncertainties on these scale factors are included as nuisance parameters.
As small differences in lepton isolation requirement performance have been observed related to the total number of jets per event, an additional uncertainty of 1.0 (0.5)\% is added to the electron (muon) ID scale factor uncertainties to account for the difference in average jet multiplicity between \Zjets and \ttbar events.
Apart from the uncertainty in the integrated luminosity, the uncertainties on the lepton ID efficiency constitute the largest source of systematic uncertainty.

As a cross-check of these scale factors, a profile maximum likelihood fit is performed with lepton ID efficiencies integrated over the full range of reconstructed lepton kinematics and treated as free parameters in the fit.
Since the lepton \pt and \sigeta selection criteria are identical across all event categories, scale factors for these integrated efficiencies can be applied to each channel based solely on lepton multiplicity.
These scale factors can then be constrained by the combination of multiple decay channels, provided that the events are not further separated in a way that breaks the approximate independence of the lepton ID efficiency.
The likelihood fit is then able to determine overall lepton efficiency scale factors with an uncertainty of 2\%, and we observe that the scale factors from tag-and-probe measurements in \Zjets events agree within this uncertainty.
Although this cross-check ultimately determines the scale factors with less precision, it demonstrates that such scale factors can in principle be derived purely in situ, taking advantage of multiple decay channels with carefully-designed event selection.

Similarly to the lepton ID efficiencies, scale factors for single-lepton trigger efficiencies are measured with the tag-and-probe method in a \Zjets-enriched CR and applied to the simulated event samples.
The uncertainty in the trigger efficiency scale factors is less than 1\%, though some dependence on the total number of jets per event has been observed.
To account for differences in average jet multiplicity between \Zjets and \ttbar events, an additional uncertainty of 1.0 (0.5)\% is considered on single electron (muon) triggers.
The uncertainty in the dilepton trigger efficiency is negligible due to offline lepton \pt requirements that are much larger than the trigger requirements.

The impact of the JES uncertainties is estimated by varying the jet momenta within 26 different uncertainty sources, following the methods described in Ref.~\cite{CMS:JME-13-004}.
Of these, 17 are found to be nonnegligible and are included in the fit.
To verify the recent jet energy calibration of the CMS detector, global JES scale factors are derived within the analysis in a coarse binning of \pt and \abseta, using the reconstructed invariant mass distributions of hadronically decaying \PW bosons in the \ljets channels.
This cross-check is designed to be sensitive to any major disagreements in the JES between data and simulation affecting the selected events.
Fitting the shift of the reconstructed \PW boson mass yields JES factors within 1.5\% of unity, a discrepancy well-covered by the included uncertainties.

For the determination of the \PQb tagging efficiencies, multinomial probabilities are used to describe the expected number of events for both signal and background samples depending on the number of \PQb jets at generator level.
The \PQb tagging efficiency is allowed to vary as a free parameter.
This in situ approach takes advantage of constraints arising from the categorization into events with 0, 1, or 2 \PQb jets~\cite{CMS:TOP-17-001}.
The misidentification rate of light and \PQc quark jets as \PQb jets is described in a similar way. 
It is allowed to vary within 10\% from that observed in simulation, and variation beyond this level is not found to affect the measurement outcome or sensitivity.

Differences in the pileup distribution between simulation and data are corrected by reweighting the simulation using weights binned in three different variables:\ the number of reconstructed vertices, the mean energy density calculated from the tracker alone, and the mean energy density calculated from the calorimeter alone.
The average of the three weights is used for the nominal pileup distribution, and the difference from the result using only the number of reconstructed vertices is used to estimate the uncertainty.

The uncertainty in the calibration of the integrated luminosity measurement, following the procedures described in Ref.~\cite{CMS:LUM-17-003}, is estimated to be 2.1\%.
The largest contributions are from the factorization bias, which arises in the van der Meer method from the assumption that the transverse luminous area factorizes in the $x$ and $y$ coordinates, and from residual beam position deviations.
Good agreement in the absolute scale is found between the independently calibrated luminosity measurements, and the integrated luminosity measured with the HF and the silicon pixel detector agrees to a level of better than 0.8\%.
Taking additional contributions due to residual differences in the time-stability and linearity between the luminosity detectors into account, the total uncertainty in the integrated luminosity is estimated to be 2.3\%.
A cross-check of the integrated luminosity using the yield of reconstructed \PZ bosons decaying into pairs of muons~\cite{CMS:DP-2023-003}, corrected for efficiencies and normalized to the fiducial cross section prediction calculated at NNLO with next-to-NNLL corrections applied, shows good agreement as well.
The integrated luminosity uncertainty is not included in the fit, but treated as an external uncertainty and added in quadrature afterwards.
The impact of varying the normalization of the backgrounds estimated from simulation by the integrated luminosity uncertainty was found to be negligible.

The uncertainty related to higher-order terms in the ME calculation is modeled by varying the renormalization and factorization scales, indicated as \muR and \muF, in the generator.
These are varied by factors of two up and down, independently and simultaneously, but avoiding cases where $\muR/\muF=0.25$ or 4~\cite{EXT:TtbarSystematic-2004}.
In order to remove effects on the cross sections of signal and background processes while still considering possible acceptance effects, each variation is normalized to preserve the respective nominal sample cross section before any event selection.
The ME scale variations are performed for the \ttbar, single-\PQt, \Zjets, and \Wjets samples, and treated as uncorrelated between the processes.

The uncertainty due to the matching of the ME to the PS simulation is estimated by varying the \hdamp parameter in \POWHEG, as described in Ref.~\cite{CMS:TOP-16-021}.
The impact of the PS scale uncertainty is estimated by independently varying the initial- and final-state radiation scales by a factor of two up and down.

The PDF uncertainty is estimated by reweighting the simulated samples to match 100 different PDF replicas in the NNPDF3.1 set, following the PDF4LHC recommendations~\cite{Butterworth:2015oua}.
The uncertainty from the choice of the value of the strong coupling constant \alpS is estimated by an analogous reweighting.
Similarly to the ME uncertainties, these variations are normalized to match the respective nominal sample content before event selection.
They are then summed in quadrature to create a single uncertainty shape template.

It has been shown in differential measurements of \sigtt at \sqrts[13] that the \pt distribution of the top quark is softer than predicted by the \POWHEG simulation used for our \ttbar signal sample~\cite{CMS:TOP-17-014, CMS:TOP-16-007, CMS:TOP-16-008}.
Because of this, an additional uncertainty is estimated by reweighting the top quark \pt of the simulation to a fixed-order calculation at NNLO in QCD and electroweak contributions~\cite{Czakon:2017wor}.
The resulting nuisance parameter is one-sided by definition.

Normalization uncertainties are included in the background estimates to account for uncertainty in the cross section values, and for effects of systematic uncertainties whose shape templates are prone to large statistical fluctuations due to low event counts.
Uncertainties of 30\% are assigned to the \Zjets, \Wjets, diboson, and QCD multijet backgrounds, all of which have little impact on the fit, and of 15\% to the \tW and $t$-channel single-\PQt backgrounds, based on the precision of experimental measurements~\cite{CMS:SMP-14-020, CMS:TOP-17-011, CMS:SMP-18-004, CMS:TOP-20-002, CMS:SMP-20-014, CMS:SMP-20-015, CMS:TOP-21-010} and consistent with the treatment in previous \ttbar measurements~\cite{CMS:TOP-15-003, CMS:TOP-16-005, CMS:TOP-17-001}.
The normalization uncertainty on the QCD multijet background is treated independently for events with nonprompt electrons and those with nonprompt muons, as the two distributions are derived by comparison with distinct sideband CRs.
In addition, bin-by-bin statistical uncertainties are applied to the signal and background processes using the method given in Ref.~\cite{Barlow:1993dm}.
These are minuscule for the simulated backgrounds but not negligible for the nonprompt background since it is estimated from data CRs with limited event counts.

To evaluate the approximate effect of a given source of uncertainty on the overall measurement precision, the fit (described in Section~\ref{sec:fit}) is repeated with the associated nuisance parameter(s) frozen at the best-fit value(s).
The total resulting fit uncertainty is then subtracted in quadrature from that of the full measurement, yielding an approximate value for the uncertainty contribution of the individual source.
In the special case of statistical uncertainty, the fit is repeated with all nuisance parameters frozen at their best-fit values and the resulting uncertainty is taken to be that arising purely from statistical effects.
The information from these fits is summarized in Table~\ref{tab:sys}.

\begin{table}[!htb]
\centering\renewcommand\arraystretch{1.1}
\topcaption{%
    Summary of the sources of uncertainty in the \sigtt measurement.
    The relative uncertainty values are approximate and given without their correlations.
    The statistical uncertainty includes contributions from both the signal and control regions.
    The combined uncertainty includes correlations between sources.
    The integrated luminosity uncertainty is listed separately.
}
\begin{tabular}{l@{}c}
    Source & Uncertainty (\%) \\
    \hline
    Lepton ID efficiencies & 1.6 \\
    Trigger efficiency & 0.3 \\
    JES & 0.6 \\
    \PQb tagging efficiency & 1.1 \\
    Pileup reweighting & 0.5 \\
    ME scale, \ttbar & 0.5 \\
    ME scale, backgrounds & 0.2 \\
    ME/PS matching & 0.1 \\
    PS scales & 0.3 \\
    PDF and \alpS & 0.3 \\
    Top quark \pt & 0.5 \\
    \tW background & 0.7 \\
    $t$-channel single-\PQt background & 0.4 \\
    \Zjets background & 0.3 \\
    \Wjets background & \makebox[0pt][r]{$<$}0.1 \\
    Diboson background & 0.6 \\
    QCD multijet background & 0.3 \\
    Statistical uncertainty & 0.5 \\ \hline
    Combined uncertainty & 2.5 \\ \hline
    Integrated luminosity & 2.3 \\
\end{tabular}
\label{tab:sys}
\end{table}

\section{Fit procedure and results}
\label{sec:fit}

\begin{figure}[!tp]
\centering
\includegraphics[width=0.9\textwidth]{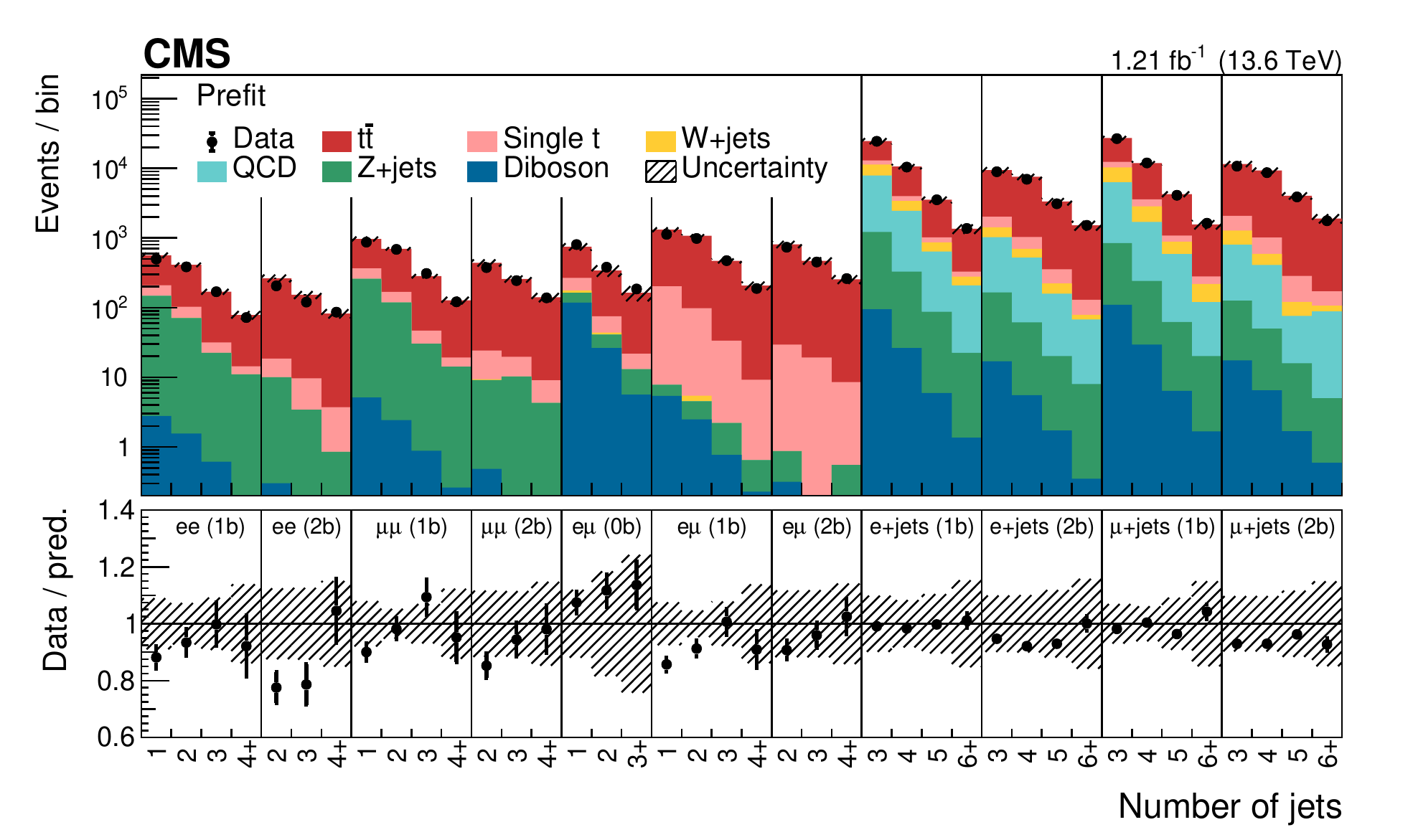}
\includegraphics[width=0.9\textwidth]{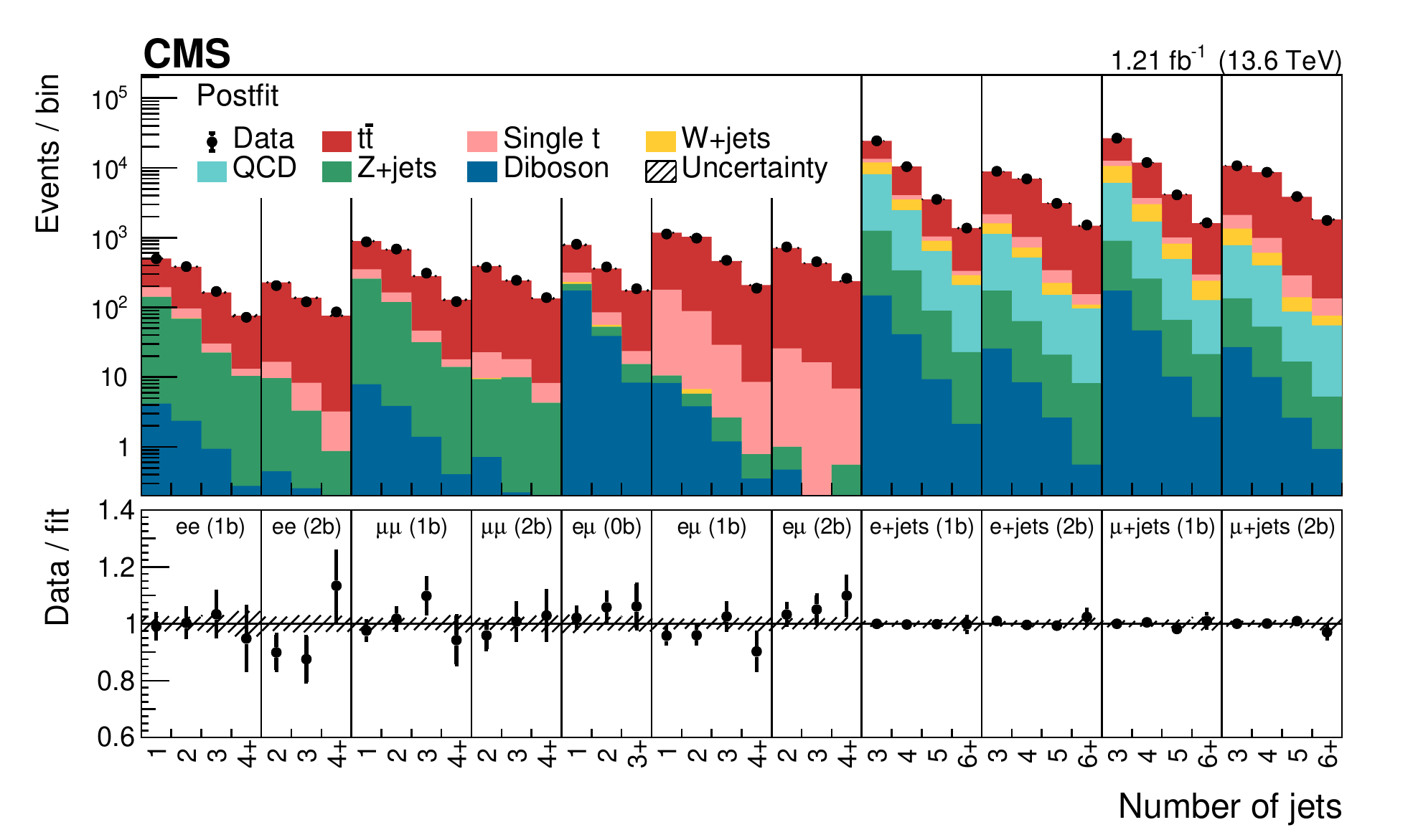}
\caption{%
    Comparison of the number of observed (points) and predicted (filled histograms) events in the final analysis binning.
    The predictions are shown before (upper) and after (lower) fitting the model to the data.
    The lower panel of each plot displays the ratio of the event yields in data to the sum of predicted signal and background yields.
    The vertical bars on the points represent the statistical uncertainties in the data, while the hatched bands represent the systematic uncertainty in the predictions, excluding the integrated luminosity.
    No \PQb jet efficiency scale factors are applied in the upper plot, and no systematic uncertainty entering into the hatched bands is intended to cover these factors, which are free parameters in the fit.
}
\label{fig:fit}
\end{figure}

In addition to the categorization by the lepton flavor and number, as well as the \PQb jet multiplicity, events are further binned by the number of jets.
The agreement in these bins, after performing the final event selection and background estimation, is shown in the upper plot of Fig.~\ref{fig:fit}.

A profile maximum likelihood fit is performed to determine the best values of \sigtt and of the nuisance parameters, and to assign the measured uncertainty, following the procedure described in Section 3.2 of Ref.~\cite{CMS:HIG-14-009}.
In the fit, statistical fluctuations follow a Poisson distribution, while uncertainties affecting solely the normalization of the samples are modeled with log-normal distributions.
For all other nuisance parameters, shape templates are generated to model the effect on the expected bin content when each parameter is varied by one standard deviation from the best estimated nominal value.
Each such parameter is assigned a term in the likelihood function following a Gaussian distribution, while a morphing function interpolates the discrete shape templates to define a smooth mapping from the value of the Gaussian random variable to the expected bin content.

The result of the fit is shown in the lower plot of Fig.~\ref{fig:fit}.
Considering only the statistical uncertainty in the data, the agreement between the data and prediction in the final analysis binning is greatly improved.
An important contribution to that effect is the estimation of the \PQb tagging efficiency in the fit.
The uncertainties are greatly reduced due to additional constraints of the nuisance parameters as well as correlations between them.

\begin{figure}[!ht]
\centering
\includegraphics[width=\textwidth]{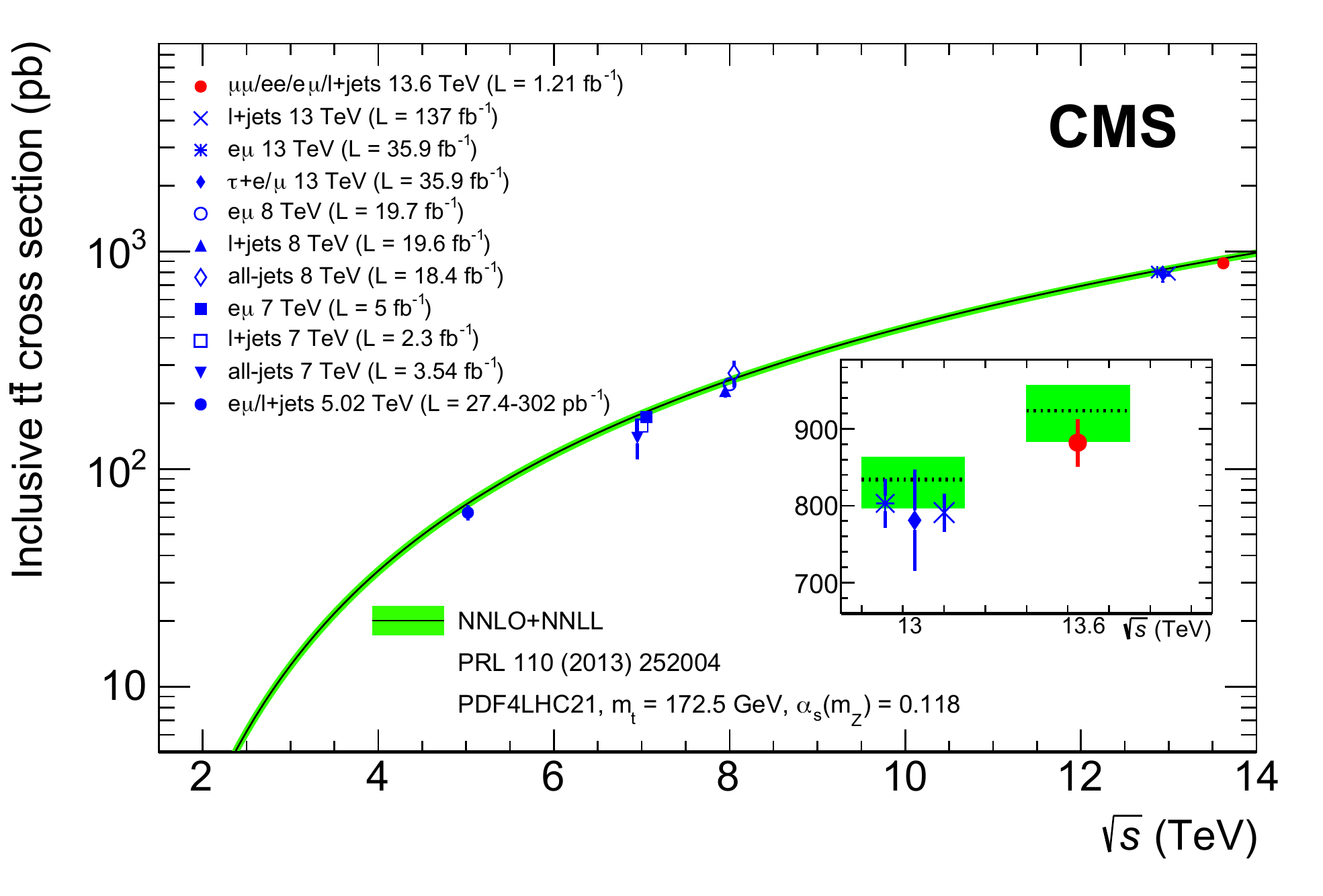}
\caption{%
    The \ttbar cross section as a function of \sqs, as obtained in this analysis (red filled circle) and in previous measurements by the CMS
    experiment~\cite{CMS:TOP-11-007, CMS:TOP-14-018, CMS:TOP-12-006, CMS:TOP-13-004, CMS:TOP-17-001, CMS:TOP-18-005, CMS:TOP-20-001, CMS:TOP-20-004} (blue markers), with vertical bars on the markers indicating the total uncertainty in the measurements.
    Points corresponding to measurements at the same \sqs are horizontally shifted for better visibility.
    The SM prediction at NNLO+NNLL precision~\cite{Czakon:2013goa} using the PDF4LHC21 PDF set~\cite{PDF4LHCWorkingGroup:2022cjn} and values of $\mtop=172.5\GeV$ and $\alpS(\mZ)=0.118$ is shown with a black line and green uncertainty bands.
    An enlarged inset is included to highlight the difference between 13 and 13.6\TeV predictions and results.
}
\label{fig:xsec}
\end{figure}

The inclusive \ttbar cross section is measured to be \xsecmeas.
This result is in good agreement with the SM prediction of \xsecpred, where the quoted uncertainty accounts for scale variations and the PDF choice.
In Fig.~\ref{fig:xsec}, the result from this measurement, with the integrated luminosity uncertainty added in quadrature to the combined uncertainty from the fit, is shown together with several measurements performed at other center-of-mass energies, along with a global comparison to the SM prediction.

The \ttbar signal simulation used in the measurement was generated using $\mtop=172.5\GeV$.
To assess the dependence of the measured cross section on the mass used in the simulation through acceptance effects, the fit is repeated for different simulation samples with masses differing by $\pm$3\GeV.
We find that for an increase (decrease) of \mtop by its current experimental uncertainty of 0.3\GeV~\cite{EXT:PDG-2022}, the measured \ttbar cross section decreases (increases) by 0.3\%.

An independent cross section measurement is performed using an event-counting method restricted to events containing an opposite-sign \emu pair and at least two jets.
This strategy follows closely the methods of Ref.~\cite{CMS:TOP-15-003}, and can also be compared to event counting measurements found in Refs.~\cite{CMS:TOP-13-004, CMS:TOP-16-005, CMS:TOP-17-001}.
With this alternative approach, the cross section is measured to be $888\pm 34\statsyst\pm 20\lum\pb$.
While the two approaches share event selection and lepton ID scale factors, the latter approach is completely independent of the \PQb tagging performance and does not use information from the jet multiplicity distribution beyond the initial selection requirement of at least two selected jets.

\section{Summary}

The first measurement of the top quark pair (\ttbar) production cross section in proton-proton collisions at \sqrts is presented.
Data recorded with the CMS detector in Summer 2022, corresponding to an integrated luminosity of \intlumi, are analyzed.
Events are selected with one or two charged leptons (electrons or muons) and additional jets.
A profile maximum likelihood fit is performed on categories defined by the number and flavors of the leptons, the total number of jets, and the number of jets identified as originating from \PQb quarks.
The fit is used to constrain the uncertainties in the \PQb tagging efficiencies and lepton selection efficiencies.
Novel cross-checks are performed on the selected \ttbar data sample to verify the lepton selection efficiencies, as well as the jet energy scale, while the cross section result itself is verified by an independent event counting approach in the \emu channel.
An inclusive \ttbar production cross section of \xsecmeas is measured, in agreement with the standard model prediction of \xsecpred.

\begin{acknowledgments}
    We congratulate our colleagues in the CERN accelerator departments for the excellent performance of the LHC and thank the technical and administrative staffs at CERN and at other CMS institutes for their contributions to the success of the CMS effort. In addition, we gratefully acknowledge the computing centres and personnel of the Worldwide LHC Computing Grid and other centres for delivering so effectively the computing infrastructure essential to our analyses. Finally, we acknowledge the enduring support for the construction and operation of the LHC, the CMS detector, and the supporting computing infrastructure provided by the following funding agencies: BMBWF and FWF (Austria); FNRS and FWO (Belgium); CNPq, CAPES, FAPERJ, FAPERGS, and FAPESP (Brazil); MES and BNSF (Bulgaria); CERN; CAS, MoST, and NSFC (China); MINCIENCIAS (Colombia); MSES and CSF (Croatia); RIF (Cyprus); SENESCYT (Ecuador); MoER, ERC PUT and ERDF (Estonia); Academy of Finland, MEC, and HIP (Finland); CEA and CNRS/IN2P3 (France); BMBF, DFG, and HGF (Germany); GSRI (Greece); NKFIH (Hungary); DAE and DST (India); IPM (Iran); SFI (Ireland); INFN (Italy); MSIP and NRF (Republic of Korea); MES (Latvia); LAS (Lithuania); MOE and UM (Malaysia); BUAP, CINVESTAV, CONACYT, LNS, SEP, and UASLP-FAI (Mexico); MOS (Montenegro); MBIE (New Zealand); PAEC (Pakistan); MES and NSC (Poland); FCT (Portugal); MESTD (Serbia); MCIN/AEI and PCTI (Spain); MOSTR (Sri Lanka); Swiss Funding Agencies (Switzerland); MST (Taipei); MHESI and NSTDA (Thailand); TUBITAK and TENMAK (Turkey); NASU (Ukraine); STFC (United Kingdom); DOE and NSF (USA).

    \hyphenation{Rachada-pisek} Individuals have received support from the Marie-Curie programme and the European Research Council and Horizon 2020 Grant, contract Nos.\ 675440, 724704, 752730, 758316, 765710, 824093, 884104, and COST Action CA16108 (European Union); the Leventis Foundation; the Alfred P.\ Sloan Foundation; the Alexander von Humboldt Foundation; the Belgian Federal Science Policy Office; the Fonds pour la Formation \`a la Recherche dans l'Industrie et dans l'Agriculture (FRIA-Belgium); the Agentschap voor Innovatie door Wetenschap en Technologie (IWT-Belgium); the F.R.S.-FNRS and FWO (Belgium) under the ``Excellence of Science -- EOS" -- be.h project n.\ 30820817; the Beijing Municipal Science \& Technology Commission, No. Z191100007219010; the Ministry of Education, Youth and Sports (MEYS) of the Czech Republic; the Hellenic Foundation for Research and Innovation (HFRI), Project Number 2288 (Greece); the Deutsche Forschungsgemeinschaft (DFG), under Germany's Excellence Strategy -- EXC 2121 ``Quantum Universe" -- 390833306, and under project number 400140256 - GRK2497; the Hungarian Academy of Sciences, the New National Excellence Program - \'UNKP, the NKFIH research grants K 124845, K 124850, K 128713, K 128786, K 129058, K 131991, K 133046, K 138136, K 143460, K 143477, 2020-2.2.1-ED-2021-00181, and TKP2021-NKTA-64 (Hungary); the Council of Science and Industrial Research, India; the Latvian Council of Science; the Ministry of Education and Science, project no. 2022/WK/14, and the National Science Center, contracts Opus 2021/41/B/ST2/01369 and 2021/43/B/ST2/01552 (Poland); the Funda\c{c}\~ao para a Ci\^encia e a Tecnologia, grant CEECIND/01334/2018 (Portugal); the National Priorities Research Program by Qatar National Research Fund; MCIN/AEI/10.13039/501100011033, ERDF ``a way of making Europe", and the Programa Estatal de Fomento de la Investigaci{\'o}n Cient{\'i}fica y T{\'e}cnica de Excelencia Mar\'{\i}a de Maeztu, grant MDM-2017-0765 and Programa Severo Ochoa del Principado de Asturias (Spain); the Chulalongkorn Academic into Its 2nd Century Project Advancement Project, and the National Science, Research and Innovation Fund via the Program Management Unit for Human Resources \& Institutional Development, Research and Innovation, grant B05F650021 (Thailand); the Kavli Foundation; the Nvidia Corporation; the SuperMicro Corporation; the Welch Foundation, contract C-1845; and the Weston Havens Foundation (USA).
\end{acknowledgments}

\bibliography{auto_generated}

\providecommand{\href}[2]{#2}\begingroup\raggedright\begin{thebibliography}{10}%
\makeatletter
\providecommand{\hrefCMSnoop }[0]{\@secondoftwo}%
\makeatother
\providecommand{\doi}{\texttt{doi:}\begingroup \urlstyle{tt}\Url}

\bibitem{CMS:TOP-11-007}
\hrefCMSnoop {}{{CMS Collaboration}, ``Measurement of the \ttbar production
  cross section in the all-jet final state in ${\Pp\Pp}$ collisions at
  $\sqrt{s}={7\TeV}$'',} \textit{ JHEP} \textbf{ 05} (2013) 065,
  \href{http://dx.doi.org/10.1007/JHEP05(2013)065}{\doi{10.1007/JHEP05(2013)065}},
  \href{http://www.arXiv.org/abs/1302.0508}{\texttt{arXiv:1302.0508}}.

\bibitem{ATLAS:2014nxi}
\hrefCMSnoop {}{{ATLAS Collaboration}, ``Measurement of the \ttbar production
  cross-section using ${\Pe\PGm}$ events with {\PQb}-tagged jets in \pp
  collisions at $\sqrt{s}=7$ and {8\TeV} with the {ATLAS} detector'',} \textit{
  Eur. Phys. J. C} \textbf{ 74} (2014) 3109,
  \href{http://dx.doi.org/10.1140/epjc/s10052-016-4501-2}{\doi{10.1140/epjc/s10052-016-4501-2}},
  \href{http://www.arXiv.org/abs/1406.5375}{\texttt{arXiv:1406.5375}}.
  [Addendum: \DOI{10.1140/epjc/s10052-016-4501-2}].

\bibitem{CMS:TOP-14-018}
\hrefCMSnoop {}{{CMS Collaboration}, ``Measurement of the \ttbar production
  cross section in the all-jets final state in ${\Pp\Pp}$ collisions at
  $\sqrt{s}={8\TeV}$'',} \textit{ Eur. Phys. J. C} \textbf{ 76} (2016) 128,
  \href{http://dx.doi.org/10.1140/epjc/s10052-016-3956-5}{\doi{10.1140/epjc/s10052-016-3956-5}},
  \href{http://www.arXiv.org/abs/1509.06076}{\texttt{arXiv:1509.06076}}.

\bibitem{CMS:TOP-12-006}
\hrefCMSnoop {}{{CMS Collaboration}, ``Measurements of the \ttbar production
  cross section in lepton+jets final states in ${\Pp\Pp}$ collisions at {8\TeV}
  and ratio of 8 to {7\TeV} cross sections'',} \textit{ Eur. Phys. J. C}
  \textbf{ 77} (2017) 15,
  \href{http://dx.doi.org/10.1140/epjc/s10052-016-4504-z}{\doi{10.1140/epjc/s10052-016-4504-z}},
  \href{http://www.arXiv.org/abs/1602.09024}{\texttt{arXiv:1602.09024}}.

\bibitem{CMS:TOP-13-004}
\hrefCMSnoop {}{{CMS Collaboration}, ``Measurement of the \ttbar production
  cross section in the ${\Pe\PGm}$ channel in proton-proton collisions at
  $\sqrt{s}=7$ and {8\TeV}'',} \textit{ JHEP} \textbf{ 08} (2016) 029,
  \href{http://dx.doi.org/10.1007/JHEP08(2016)029}{\doi{10.1007/JHEP08(2016)029}},
  \href{http://www.arXiv.org/abs/1603.02303}{\texttt{arXiv:1603.02303}}.

\bibitem{ATLAS:2017jkf}
\hrefCMSnoop {}{{ATLAS Collaboration}, ``Measurement of the \ttbar production
  cross section in the {\PGt}+jets final state in \pp collisions at
  $\sqrt{s}={8\TeV}$ using the {ATLAS} detector'',} \textit{ Phys. Rev. D}
  \textbf{ 95} (2017) 072003,
  \href{http://dx.doi.org/10.1103/PhysRevD.95.072003}{\doi{10.1103/PhysRevD.95.072003}},
  \href{http://www.arXiv.org/abs/1702.08839}{\texttt{arXiv:1702.08839}}.

\bibitem{ATLAS:2017wvi}
\hrefCMSnoop {}{{ATLAS Collaboration}, ``Measurement of the inclusive and
  fiducial \ttbar production cross-sections in the lepton+jets channel in \pp
  collisions at $\sqrt{s}={8\TeV}$ with the {ATLAS} detector'',} \textit{ Eur.
  Phys. J. C} \textbf{ 78} (2018) 487,
  \href{http://dx.doi.org/10.1140/epjc/s10052-018-5904-z}{\doi{10.1140/epjc/s10052-018-5904-z}},
  \href{http://www.arXiv.org/abs/1712.06857}{\texttt{arXiv:1712.06857}}.

\bibitem{LHCb:2018usb}
\hrefCMSnoop {}{{LHCb Collaboration}, ``Measurement of forward top pair
  production in the dilepton channel in \pp collisions at
  $\sqrt{s}={13\TeV}$'',} \textit{ JHEP} \textbf{ 08} (2018) 174,
  \href{http://dx.doi.org/10.1007/JHEP08(2018)174}{\doi{10.1007/JHEP08(2018)174}},
  \href{http://www.arXiv.org/abs/1803.05188}{\texttt{arXiv:1803.05188}}.

\bibitem{CMS:TOP-17-001}
\hrefCMSnoop {}{{CMS Collaboration}, ``Measurement of the \ttbar production
  cross section, the top quark mass, and the strong coupling constant using
  dilepton events in ${\Pp\Pp}$ collisions at $\sqrt{s}={13\TeV}$'',} \textit{
  Eur. Phys. J. C} \textbf{ 79} (2019) 368,
  \href{http://dx.doi.org/10.1140/epjc/s10052-019-6863-8}{\doi{10.1140/epjc/s10052-019-6863-8}},
  \href{http://www.arXiv.org/abs/1812.10505}{\texttt{arXiv:1812.10505}}.

\bibitem{CMS:TOP-18-005}
\hrefCMSnoop {}{{CMS Collaboration}, ``Measurement of the top quark pair
  production cross section in dilepton final states containing one {\PGt}
  lepton in ${\Pp\Pp}$ collisions at $\sqrt{s}={13\TeV}$'',} \textit{ JHEP}
  \textbf{ 02} (2020) 191,
  \href{http://dx.doi.org/10.1007/JHEP02(2020)191}{\doi{10.1007/JHEP02(2020)191}},
  \href{http://www.arXiv.org/abs/1911.13204}{\texttt{arXiv:1911.13204}}.

\bibitem{ATLAS:2020ccu}
\hrefCMSnoop {}{{ATLAS Collaboration}, ``Measurements of top-quark pair single-
  and double-differential cross-sections in the all-hadronic channel in \pp
  collisions at \sqrts[13] using the {ATLAS} detector'',} \textit{ JHEP}
  \textbf{ 01} (2021) 033,
  \href{http://dx.doi.org/10.1007/JHEP01(2021)033}{\doi{10.1007/JHEP01(2021)033}},
  \href{http://www.arXiv.org/abs/2006.09274}{\texttt{arXiv:2006.09274}}.

\bibitem{ATLAS:2020aln}
\hrefCMSnoop {}{{ATLAS Collaboration}, ``Measurement of the \ttbar production
  cross-section in the lepton+jets channel at \sqrts[13] with the {ATLAS}
  experiment'',} \textit{ Phys. Lett. B} \textbf{ 810} (2020) 135797,
  \href{http://dx.doi.org/10.1016/j.physletb.2020.135797}{\doi{10.1016/j.physletb.2020.135797}},
  \href{http://www.arXiv.org/abs/2006.13076}{\texttt{arXiv:2006.13076}}.

\bibitem{CMS:TOP-20-001}
\hrefCMSnoop {}{{CMS Collaboration}, ``Measurement of differential \ttbar
  production cross sections in the full kinematic range using lepton+jets
  events from proton-proton collisions at $\sqrt{s}={13\TeV}$'',} \textit{
  Phys. Rev. D} \textbf{ 104} (2021) 092013,
  \href{http://dx.doi.org/10.1103/PhysRevD.104.092013}{\doi{10.1103/PhysRevD.104.092013}},
  \href{http://www.arXiv.org/abs/2108.02803}{\texttt{arXiv:2108.02803}}.

\bibitem{CMS:TOP-20-004}
\hrefCMSnoop {}{{CMS Collaboration}, ``Measurement of the inclusive \ttbar
  production cross section in proton-proton collisions at
  $\sqrt{s}={5.02\TeV}$'',} \textit{ JHEP} \textbf{ 04} (2022) 144,
  \href{http://dx.doi.org/10.1007/JHEP04(2022)144}{\doi{10.1007/JHEP04(2022)144}},
  \href{http://www.arXiv.org/abs/2112.09114}{\texttt{arXiv:2112.09114}}.

\bibitem{CMS:TOP-18-014}
\hrefCMSnoop {}{{ATLAS and CMS Collaborations}, ``Combination of inclusive
  top-quark pair production cross-section measurements using {ATLAS} and {CMS}
  data at $\sqrt{s}=7$ and {8\TeV}'',} \textit{ JHEP} \textbf{ 07} (2023) 213,
  \href{http://dx.doi.org/10.1007/JHEP07(2023)213}{\doi{10.1007/JHEP07(2023)213}},
  \href{http://www.arXiv.org/abs/2205.13830}{\texttt{arXiv:2205.13830}}.

\bibitem{ATLAS:2022jbj}
\hrefCMSnoop {}{{ATLAS Collaboration}, ``Measurement of the \ttbar production
  cross-section in \pp collisions at \sqrts[5.02] with the {ATLAS} detector'',}
  \textit{ JHEP} \textbf{ 06} (2023) 138,
  \href{http://dx.doi.org/10.1007/JHEP06(2023)138}{\doi{10.1007/JHEP06(2023)138}},
  \href{http://www.arXiv.org/abs/2207.01354}{\texttt{arXiv:2207.01354}}.

\bibitem{ATLAS:2022xny}
\hrefCMSnoop {}{{ATLAS Collaboration}, ``Measurement of the inclusive \ttbar
  production cross section in the lepton+jets channel in \pp collisions at
  $\sqrt{s}={7\TeV}$ with the {ATLAS} detector using support vector
  machines'',} 2022.
  \href{http://www.arXiv.org/abs/2212.00571}{\texttt{arXiv:2212.00571}}.
  Accepted by \textit{Phys. Rev. D}.

\bibitem{ATLAS:2023gsl}
\hrefCMSnoop {}{{ATLAS Collaboration}, ``Inclusive and differential
  cross-sections for dilepton \ttbar production measured in $\sqrt{s}={13\TeV}$
  \pp collisions with the {ATLAS} detector'',} \textit{ JHEP} \textbf{ 07}
  (2023) 141,
  \href{http://dx.doi.org/10.1007/JHEP07(2023)141}{\doi{10.1007/JHEP07(2023)141}},
  \href{http://www.arXiv.org/abs/2303.15340}{\texttt{arXiv:2303.15340}}.

\bibitem{Cacciari:2011hy}
M.~Cacciari\hrefCMSnoop {}{ {et~al.}, ``Top-pair production at hadron colliders
  with next-to-next-to-leading logarithmic soft-gluon resummation'',} \textit{
  Phys. Lett. B} \textbf{ 710} (2012) 612,
  \href{http://dx.doi.org/10.1016/j.physletb.2012.03.013}{\doi{10.1016/j.physletb.2012.03.013}},
  \href{http://www.arXiv.org/abs/1111.5869}{\texttt{arXiv:1111.5869}}.

\bibitem{Barnreuther:2012wtj}
\hrefCMSnoop {}{P.~B{\"a}rnreuther, M.~Czakon, and A.~Mitov, ``Percent level
  precision physics at the {Tevatron}: Next-to-next-to-leading order {QCD}
  corrections to $\qqbar\to\ttbar+{\PX}$'',} \textit{ Phys. Rev. Lett.}
  \textbf{ 109} (2012) 132001,
  \href{http://dx.doi.org/10.1103/PhysRevLett.109.132001}{\doi{10.1103/PhysRevLett.109.132001}},
  \href{http://www.arXiv.org/abs/1204.5201}{\texttt{arXiv:1204.5201}}.

\bibitem{Czakon:2012zr}
\hrefCMSnoop {}{M.~Czakon and A.~Mitov, ``{NNLO} corrections to top-pair
  production at hadron colliders: the all-fermionic scattering channels'',}
  \textit{ JHEP} \textbf{ 12} (2012) 054,
  \href{http://dx.doi.org/10.1007/JHEP12(2012)054}{\doi{10.1007/JHEP12(2012)054}},
  \href{http://www.arXiv.org/abs/1207.0236}{\texttt{arXiv:1207.0236}}.

\bibitem{Czakon:2012pz}
\hrefCMSnoop {}{M.~Czakon and A.~Mitov, ``{NNLO} corrections to top pair
  production at hadron colliders: the quark-gluon reaction'',} \textit{ JHEP}
  \textbf{ 01} (2013) 080,
  \href{http://dx.doi.org/10.1007/JHEP01(2013)080}{\doi{10.1007/JHEP01(2013)080}},
  \href{http://www.arXiv.org/abs/1210.6832}{\texttt{arXiv:1210.6832}}.

\bibitem{Czakon:2013goa}
\hrefCMSnoop {}{M.~Czakon, P.~Fiedler, and A.~Mitov, ``Total top-quark
  pair-production cross section at hadron colliders through
  $\mathcal{O}({\alpS}^4)$'',} \textit{ Phys. Rev. Lett.} \textbf{ 110} (2013)
  252004,
  \href{http://dx.doi.org/10.1103/PhysRevLett.110.252004}{\doi{10.1103/PhysRevLett.110.252004}},
  \href{http://www.arXiv.org/abs/1303.6254}{\texttt{arXiv:1303.6254}}.

\bibitem{Catani:2019iny}
S.~Catani\hrefCMSnoop {}{ {et~al.}, ``Top-quark pair hadroproduction at
  next-to-next-to-leading order in {QCD}'',} \textit{ Phys. Rev. D} \textbf{
  99} (2019) 051501,
  \href{http://dx.doi.org/10.1103/PhysRevD.99.051501}{\doi{10.1103/PhysRevD.99.051501}},
  \href{http://www.arXiv.org/abs/1901.04005}{\texttt{arXiv:1901.04005}}.

\bibitem{Czakon:2011xx}
\hrefCMSnoop {}{M.~Czakon and A.~Mitov, ``\textsc{top++}: a program for the
  calculation of the top-pair cross-section at hadron colliders'',} \textit{
  Comput. Phys. Commun.} \textbf{ 185} (2014) 2930,
  \href{http://dx.doi.org/10.1016/j.cpc.2014.06.021}{\doi{10.1016/j.cpc.2014.06.021}},
  \href{http://www.arXiv.org/abs/1112.5675}{\texttt{arXiv:1112.5675}}.

\bibitem{PDF4LHCWorkingGroup:2022cjn}
\hrefCMSnoop {}{{PDF4LHC Working Group}, R.~D. Ball {et~al.}, ``The {PDF4LHC21}
  combination of global {PDF} fits for the {LHC} {Run~III}'',} \textit{ J.
  Phys. G} \textbf{ 49} (2022) 080501,
  \href{http://dx.doi.org/10.1088/1361-6471/ac7216}{\doi{10.1088/1361-6471/ac7216}},
  \href{http://www.arXiv.org/abs/2203.05506}{\texttt{arXiv:2203.05506}}.

\bibitem{hepdata}
\hrefCMSnoop {}{}{HEPData} record for this analysis, 2023.
\newblock
  \href{http://dx.doi.org/10.17182/hepdata.135832}{\doi{10.17182/hepdata.135832}}.

\bibitem{CMS:Detector-2008}
\hrefCMSnoop {}{{CMS Collaboration}, ``The {CMS} experiment at the {CERN}
  {LHC}'',} \textit{ JINST} \textbf{ 3} (2008) S08004,
  \href{http://dx.doi.org/10.1088/1748-0221/3/08/S08004}{\doi{10.1088/1748-0221/3/08/S08004}}.

\bibitem{CMS:TRG-17-001}
\hrefCMSnoop {}{{CMS Collaboration}, ``Performance of the {CMS} {\Lone} trigger
  in proton-proton collisions at $\sqrt{s}={13\TeV}$'',} \textit{ JINST}
  \textbf{ 15} (2020) P10017,
  \href{http://dx.doi.org/10.1088/1748-0221/15/10/P10017}{\doi{10.1088/1748-0221/15/10/P10017}},
  \href{http://www.arXiv.org/abs/2006.10165}{\texttt{arXiv:2006.10165}}.

\bibitem{CMS:TRG-12-001}
\hrefCMSnoop {}{{CMS Collaboration}, ``The {CMS} trigger system'',} \textit{
  JINST} \textbf{ 12} (2017) P01020,
  \href{http://dx.doi.org/10.1088/1748-0221/12/01/P01020}{\doi{10.1088/1748-0221/12/01/P01020}},
  \href{http://www.arXiv.org/abs/1609.02366}{\texttt{arXiv:1609.02366}}.

\bibitem{CMS:TDR-010}
\href {https://cds.cern.ch/record/1481837}{{CMS Collaboration}, ``{CMS}
  technical design report for the {Phase 1} upgrade of the hadron
  calorimeter'',} CMS Technical Proposal CERN-LHCC-2012-015, CMS-TDR-010, 2012.

\bibitem{CMS:DP-2022-047}
\href {https://cds.cern.ch/record/2839741}{{CMS Collaboration}, ``Performance
  of the {CMS} phase-1 pixel detector with {Run~3} data'',} CMS Detector
  Performance Note CMS-DP-2022-047, 2022.

\bibitem{CMS:DP-2023-004}
\href {https://cds.cern.ch/record/2851656}{{CMS Collaboration}, ``Commissioning
  {CMS} online reconstruction with {GPUs}'',} CMS Detector Performance Note
  CMS-DP-2023-004, 2023.

\bibitem{CMS:NOTE-2022-007}
\hrefCMSnoop {}{{CMS BRIL} Collaboration, ``The {Pixel Luminosity Telescope}: a
  detector for luminosity measurement at {CMS} using silicon pixel sensors'',}
  \textit{ Eur. Phys. J. C} \textbf{ 83} (2023) 673,
  \href{http://dx.doi.org/10.1140/epjc/s10052-023-11713-6}{\doi{10.1140/epjc/s10052-023-11713-6}},
  \href{http://www.arXiv.org/abs/2206.08870}{\texttt{arXiv:2206.08870}}.

\bibitem{Karunarathna:2022tyd}
\hrefCMSnoop {}{{CMS Collaboration}, ``Run 3 luminosity measurements with the
  {Pixel Luminosity Telescope}'',} in \textit{ {Proc. 41st Int. Conf. on High
  Energy Phys. (ICHEP 2022): Bologna, Italy, July 6--13, 2022}}.
\newblock 2022.
\newblock [PoS (ICHEP2022) 936].
  \href{http://dx.doi.org/10.22323/1.414.0936}{\doi{10.22323/1.414.0936}}.

\bibitem{Wanczyk:2022avs}
\hrefCMSnoop {}{{CMS Collaboration}, ``Upgraded {CMS Fast Beam Condition
  Monitor} for {LHC Run} 3 online luminosity and beam-induced background
  measurements'',} in \textit{ {Proc. 11th Int. Beam Instrumentation Conf.
  (IBIC 2022): Krakow, Poland, September 11--15, 2022}}.
\newblock 2022.
\newblock [JACoW (IBIC2022) 540].
  \href{http://dx.doi.org/10.18429/JACoW-IBIC2022-TH2C2}{\doi{10.18429/JACoW-IBIC2022-TH2C2}}.

\bibitem{CMS:PRF-14-001}
\hrefCMSnoop {}{{CMS Collaboration}, ``Particle-flow reconstruction and global
  event description with the {CMS} detector'',} \textit{ JINST} \textbf{ 12}
  (2017) P10003,
  \href{http://dx.doi.org/10.1088/1748-0221/12/10/P10003}{\doi{10.1088/1748-0221/12/10/P10003}},
  \href{http://www.arXiv.org/abs/1706.04965}{\texttt{arXiv:1706.04965}}.

\bibitem{CMS:TDR-15-02}
\href {https://cds.cern.ch/record/2020886}{{CMS Collaboration}, ``Technical
  proposal for the {Phase-II} upgrade of the {Compact Muon Solenoid}'',} CMS
  Technical Proposal CERN-LHCC-2015-010, CMS-TDR-15-02, 2015.

\bibitem{EXT:AK-2008}
\hrefCMSnoop {}{M.~Cacciari, G.~P. Salam, and G.~Soyez, ``The anti-\kt jet
  clustering algorithm'',} \textit{ JHEP} \textbf{ 04} (2008) 063,
  \href{http://dx.doi.org/10.1088/1126-6708/2008/04/063}{\doi{10.1088/1126-6708/2008/04/063}},
  \href{http://www.arXiv.org/abs/0802.1189}{\texttt{arXiv:0802.1189}}.

\bibitem{EXT:FastJet-2012}
\hrefCMSnoop {}{M.~Cacciari, G.~P. Salam, and G.~Soyez, ``{\FASTJET} user
  manual'',} \textit{ Eur. Phys. J. C} \textbf{ 72} (2012) 1896,
  \href{http://dx.doi.org/10.1140/epjc/s10052-012-1896-2}{\doi{10.1140/epjc/s10052-012-1896-2}},
  \href{http://www.arXiv.org/abs/1111.6097}{\texttt{arXiv:1111.6097}}.

\bibitem{CMS:JME-13-004}
\hrefCMSnoop {}{{CMS Collaboration}, ``Jet energy scale and resolution in the
  {CMS} experiment in ${\Pp\Pp}$ collisions at {8\TeV}'',} \textit{ JINST}
  \textbf{ 12} (2017) P02014,
  \href{http://dx.doi.org/10.1088/1748-0221/12/02/P02014}{\doi{10.1088/1748-0221/12/02/P02014}},
  \href{http://www.arXiv.org/abs/1607.03663}{\texttt{arXiv:1607.03663}}.

\bibitem{CMS:JME-18-001}
\hrefCMSnoop {}{{CMS Collaboration}, ``Pileup mitigation at {CMS} in {13\TeV}
  data'',} \textit{ JINST} \textbf{ 15} (2020) P09018,
  \href{http://dx.doi.org/10.1088/1748-0221/15/09/P09018}{\doi{10.1088/1748-0221/15/09/P09018}},
  \href{http://www.arXiv.org/abs/2003.00503}{\texttt{arXiv:2003.00503}}.

\bibitem{EXT:PUPPI-2014}
\hrefCMSnoop {}{D.~Bertolini, P.~Harris, M.~Low, and N.~Tran, ``Pileup per
  particle identification'',} \textit{ JHEP} \textbf{ 10} (2014) 059,
  \href{http://dx.doi.org/10.1007/JHEP10(2014)059}{\doi{10.1007/JHEP10(2014)059}},
  \href{http://www.arXiv.org/abs/1407.6013}{\texttt{arXiv:1407.6013}}.

\bibitem{CMS:DP-2022-054}
\href {https://cds.cern.ch/record/2841534}{{CMS Collaboration}, ``Jet energy
  scale and resolution measurements using prompt {Run~3} data collected by
  {CMS} in the first months of 2022 at {13.6\TeV}'',} CMS Detector Performance
  Note CMS-DP-2022-054, 2022.

\bibitem{CMS:JME-16-003}
\href {https://cds.cern.ch/record/2256875}{{CMS Collaboration}, ``Jet
  algorithms performance in {13\TeV} data'',} CMS Physics Analysis Summary
  CMS-PAS-JME-16-003, 2017.

\bibitem{CMS:EWK-10-002}
\hrefCMSnoop {}{{CMS Collaboration}, ``Measurements of inclusive {\PW} and
  {\PZ} cross sections in ${\Pp\Pp}$ collisions at $\sqrt{s}={7\TeV}$'',}
  \textit{ JHEP} \textbf{ 01} (2011) 080,
  \href{http://dx.doi.org/10.1007/JHEP01(2011)080}{\doi{10.1007/JHEP01(2011)080}},
  \href{http://www.arXiv.org/abs/1012.2466}{\texttt{arXiv:1012.2466}}.

\bibitem{vanderMeer:296752}
\href {https://cds.cern.ch/record/296752}{S.~van~der Meer, ``Calibration of the
  effective beam height in the {ISR}'',} ISR Report CERN-ISR-PO-68-31, 1968.

\bibitem{Grafstrom:2015foa}
\hrefCMSnoop {}{P.~Grafstr{\"o}m and W.~Kozanecki, ``Luminosity determination
  at proton colliders'',} \textit{ Prog. Part. Nucl. Phys.} \textbf{ 81} (2015)
  97,
  \href{http://dx.doi.org/10.1016/j.ppnp.2014.11.002}{\doi{10.1016/j.ppnp.2014.11.002}}.

\bibitem{CMS:LUM-17-003}
\hrefCMSnoop {}{{CMS Collaboration}, ``Precision luminosity measurement in
  proton-proton collisions at $\sqrt{s}={13\TeV}$ in 2015 and 2016 at {CMS}'',}
  \textit{ Eur. Phys. J. C} \textbf{ 81} (2021) 800,
  \href{http://dx.doi.org/10.1140/epjc/s10052-021-09538-2}{\doi{10.1140/epjc/s10052-021-09538-2}},
  \href{http://www.arXiv.org/abs/2104.01927}{\texttt{arXiv:2104.01927}}.

\bibitem{EXT:Powheg-ttbar-2007}
\hrefCMSnoop {}{S.~Frixione, G.~Ridolfi, and P.~Nason, ``A positive-weight
  next-to-leading-order {Monte Carlo} for heavy flavour hadroproduction'',}
  \textit{ JHEP} \textbf{ 09} (2007) 126,
  \href{http://dx.doi.org/10.1088/1126-6708/2007/09/126}{\doi{10.1088/1126-6708/2007/09/126}},
  \href{http://www.arXiv.org/abs/0707.3088}{\texttt{arXiv:0707.3088}}.

\bibitem{EXT:Powheg-2007}
\hrefCMSnoop {}{S.~Frixione, P.~Nason, and C.~Oleari, ``Matching {NLO} {QCD}
  computations with parton shower simulations: the {\POWHEG} method'',}
  \textit{ JHEP} \textbf{ 11} (2007) 070,
  \href{http://dx.doi.org/10.1088/1126-6708/2007/11/070}{\doi{10.1088/1126-6708/2007/11/070}},
  \href{http://www.arXiv.org/abs/0709.2092}{\texttt{arXiv:0709.2092}}.

\bibitem{EXT:Powheg-2010}
\hrefCMSnoop {}{S.~Alioli, P.~Nason, C.~Oleari, and E.~Re, ``A general
  framework for implementing {NLO} calculations in shower {Monte Carlo}
  programs: the {\POWHEG} \textsc{box}'',} \textit{ JHEP} \textbf{ 06} (2010)
  043,
  \href{http://dx.doi.org/10.1007/JHEP06(2010)043}{\doi{10.1007/JHEP06(2010)043}},
  \href{http://www.arXiv.org/abs/1002.2581}{\texttt{arXiv:1002.2581}}.

\bibitem{EXT:Madspin-2013}
\hrefCMSnoop {}{P.~Artoisenet, R.~Frederix, O.~Mattelaer, and R.~Rietkerk,
  ``Automatic spin-entangled decays of heavy resonances in {Monte Carlo}
  simulations'',} \textit{ JHEP} \textbf{ 03} (2013) 015,
  \href{http://dx.doi.org/10.1007/JHEP03(2013)015}{\doi{10.1007/JHEP03(2013)015}},
  \href{http://www.arXiv.org/abs/1212.3460}{\texttt{arXiv:1212.3460}}.

\bibitem{EXT:MG5aMCatNLO-2014}
J.~Alwall\hrefCMSnoop {}{ {et~al.}, ``The automated computation of tree-level
  and next-to-leading order differential cross sections, and their matching to
  parton shower simulations'',} \textit{ JHEP} \textbf{ 07} (2014) 079,
  \href{http://dx.doi.org/10.1007/JHEP07(2014)079}{\doi{10.1007/JHEP07(2014)079}},
  \href{http://www.arXiv.org/abs/1405.0301}{\texttt{arXiv:1405.0301}}.

\bibitem{EXT:Pythia-2015}
T.~Sj{\"o}strand\hrefCMSnoop {}{ {et~al.}, ``An introduction to
  {\PYTHIA8.2}'',} \textit{ Comput. Phys. Commun.} \textbf{ 191} (2015) 159,
  \href{http://dx.doi.org/10.1016/j.cpc.2015.01.024}{\doi{10.1016/j.cpc.2015.01.024}},
  \href{http://www.arXiv.org/abs/1410.3012}{\texttt{arXiv:1410.3012}}.

\bibitem{EXT:NNPDF-2017}
\hrefCMSnoop {}{{NNPDF} Collaboration, ``Parton distributions from
  high-precision collider data'',} \textit{ Eur. Phys. J. C} \textbf{ 77}
  (2017) 663,
  \href{http://dx.doi.org/10.1140/epjc/s10052-017-5199-5}{\doi{10.1140/epjc/s10052-017-5199-5}},
  \href{http://www.arXiv.org/abs/1706.00428}{\texttt{arXiv:1706.00428}}.

\bibitem{CMS:GEN-17-001}
\hrefCMSnoop {}{{CMS Collaboration}, ``Extraction and validation of a new set
  of {CMS} {\PYTHIA8} tunes from underlying-event measurements'',} \textit{
  Eur. Phys. J. C} \textbf{ 80} (2020) 4,
  \href{http://dx.doi.org/10.1140/epjc/s10052-019-7499-4}{\doi{10.1140/epjc/s10052-019-7499-4}},
  \href{http://www.arXiv.org/abs/1903.12179}{\texttt{arXiv:1903.12179}}.

\bibitem{EXT:Merging-2008}
J.~Alwall\hrefCMSnoop {}{ {et~al.}, ``Comparative study of various algorithms
  for the merging of parton showers and matrix elements in hadronic
  collisions'',} \textit{ Eur. Phys. J. C} \textbf{ 53} (2008) 473,
  \href{http://dx.doi.org/10.1140/epjc/s10052-007-0490-5}{\doi{10.1140/epjc/s10052-007-0490-5}},
  \href{http://www.arXiv.org/abs/0706.2569}{\texttt{arXiv:0706.2569}}.

\bibitem{EXT:GEANT4-2002}
\hrefCMSnoop {}{{GEANT4} Collaboration, ``{\GEANTfour}---a simulation
  toolkit'',} \textit{ Nucl. Instrum. Meth. A} \textbf{ 506} (2003) 250,
  \href{http://dx.doi.org/10.1016/S0168-9002(03)01368-8}{\doi{10.1016/S0168-9002(03)01368-8}}.

\bibitem{Campbell:2020fhf}
\hrefCMSnoop {}{J.~Campbell, T.~Neumann, and Z.~Sullivan, ``Single-top-quark
  production in the $t$-channel at {NNLO}'',} \textit{ JHEP} \textbf{ 02}
  (2021) 040,
  \href{http://dx.doi.org/10.1007/JHEP02(2021)040}{\doi{10.1007/JHEP02(2021)040}},
  \href{http://www.arXiv.org/abs/2012.01574}{\texttt{arXiv:2012.01574}}.

\bibitem{Kidonakis:2021vob}
\hrefCMSnoop {}{N.~Kidonakis and N.~Yamanaka, ``Higher-order corrections for
  {\tW} production at high-energy hadron colliders'',} \textit{ JHEP} \textbf{
  05} (2021) 278,
  \href{http://dx.doi.org/10.1007/JHEP05(2021)278}{\doi{10.1007/JHEP05(2021)278}},
  \href{http://www.arXiv.org/abs/2102.11300}{\texttt{arXiv:2102.11300}}.

\bibitem{Camarda:2019zyx}
S.~Camarda\hrefCMSnoop {}{ {et~al.}, ``\textsc{DYTurbo}: Fast predictions for
  {Drell--Yan} processes'',} \textit{ Eur. Phys. J. C} \textbf{ 80} (2020) 251,
  \href{http://dx.doi.org/10.1140/epjc/s10052-020-7757-5}{\doi{10.1140/epjc/s10052-020-7757-5}},
  \href{http://www.arXiv.org/abs/1910.07049}{\texttt{arXiv:1910.07049}}.
  [Erratum: \DOI{10.1140/epjc/s10052-020-7972-0}].

\bibitem{Grazzini:2017mhc}
\hrefCMSnoop {}{M.~Grazzini, S.~Kallweit, and M.~Wiesemann, ``Fully
  differential {NNLO} computations with \textsc{matrix}'',} \textit{ Eur. Phys.
  J. C} \textbf{ 78} (2018) 537,
  \href{http://dx.doi.org/10.1140/epjc/s10052-018-5771-7}{\doi{10.1140/epjc/s10052-018-5771-7}},
  \href{http://www.arXiv.org/abs/1711.06631}{\texttt{arXiv:1711.06631}}.

\bibitem{CMS:EGM-17-001}
\hrefCMSnoop {}{{CMS Collaboration}, ``Electron and photon reconstruction and
  identification with the {CMS} experiment at the {CERN} {LHC}'',} \textit{
  JINST} \textbf{ 16} (2021) P05014,
  \href{http://dx.doi.org/10.1088/1748-0221/16/05/P05014}{\doi{10.1088/1748-0221/16/05/P05014}},
  \href{http://www.arXiv.org/abs/2012.06888}{\texttt{arXiv:2012.06888}}.

\bibitem{CMS:MUO-16-001}
\hrefCMSnoop {}{{CMS Collaboration}, ``Performance of the {CMS} muon detector
  and muon reconstruction with proton-proton collisions at
  $\sqrt{s}={13\TeV}$'',} \textit{ JINST} \textbf{ 13} (2018) P06015,
  \href{http://dx.doi.org/10.1088/1748-0221/13/06/P06015}{\doi{10.1088/1748-0221/13/06/P06015}},
  \href{http://www.arXiv.org/abs/1804.04528}{\texttt{arXiv:1804.04528}}.

\bibitem{CMS:BTV-16-002}
\hrefCMSnoop {}{{CMS Collaboration}, ``Identification of heavy-flavour jets
  with the {CMS} detector in ${\Pp\Pp}$ collisions at {13\TeV}'',} \textit{
  JINST} \textbf{ 13} (2018) P05011,
  \href{http://dx.doi.org/10.1088/1748-0221/13/05/P05011}{\doi{10.1088/1748-0221/13/05/P05011}},
  \href{http://www.arXiv.org/abs/1712.07158}{\texttt{arXiv:1712.07158}}.

\bibitem{CMS:DP-2018-058}
\href {https://cds.cern.ch/record/2646773}{{CMS Collaboration}, ``Performance
  of the {DeepJet} {\PQb} tagging algorithm using 41.9\fbinv of data from
  proton-proton collisions at {13\TeV} with {Phase} 1 {CMS} detector'',} CMS
  Detector Performance Note CMS-DP-2018-058, 2018.

\bibitem{EXT:DeepJet-2020}
E.~Bols\hrefCMSnoop {}{ {et~al.}, ``Jet flavour classification using
  {DeepJet}'',} \textit{ JINST} \textbf{ 15} (2020) P12012,
  \href{http://dx.doi.org/10.1088/1748-0221/15/12/P12012}{\doi{10.1088/1748-0221/15/12/P12012}},
  \href{http://www.arXiv.org/abs/2008.10519}{\texttt{arXiv:2008.10519}}.

\bibitem{EXT:PDG-2022}
\hrefCMSnoop {}{{Particle Data Group}, R.~L. Workman {et~al.}, ``Review of
  particle physics'',} \textit{ Prog. Theor. Exp. Phys.} \textbf{ 2022} (2022)
  083C01,
  \href{http://dx.doi.org/10.1093/ptep/ptac097}{\doi{10.1093/ptep/ptac097}}.

\bibitem{CMS:EXO-16-049}
\hrefCMSnoop {}{{CMS Collaboration}, ``Search for dark matter particles
  produced in association with a top quark pair at $\sqrt{s}={13\TeV}$'',}
  \textit{ Phys. Rev. Lett.} \textbf{ 122} (2019) 011803,
  \href{http://dx.doi.org/10.1103/PhysRevLett.122.011803}{\doi{10.1103/PhysRevLett.122.011803}},
  \href{http://www.arXiv.org/abs/1807.06522}{\texttt{arXiv:1807.06522}}.

\bibitem{CMS:DP-2023-003}
\href {https://cds.cern.ch/record/2851655}{{CMS Collaboration}, ``Luminosity
  monitoring with {\PZ} counting in early 2022 data'',} CMS Detector
  Performance Note CMS-DP-2023-003, 2023.

\bibitem{EXT:TtbarSystematic-2004}
M.~Cacciari\hrefCMSnoop {}{ {et~al.}, ``The \ttbar cross-section at 1.8 and
  {1.96\TeV}: a study of the systematics due to parton densities and scale
  dependence'',} \textit{ JHEP} \textbf{ 04} (2004) 068,
  \href{http://dx.doi.org/10.1088/1126-6708/2004/04/068}{\doi{10.1088/1126-6708/2004/04/068}},
  \href{http://www.arXiv.org/abs/hep-ph/0303085}{\texttt{arXiv:hep-ph/0303085}}.

\bibitem{CMS:TOP-16-021}
\href {https://cds.cern.ch/record/2235192}{{CMS Collaboration},
  ``Investigations of the impact of the parton shower tuning in \mbox{\PYTHIA8}
  in the modelling of \ttbar at $\sqrt{s}=8$ and {13\TeV}'',} CMS Physics
  Analysis Summary CMS-PAS-TOP-16-021, 2016.

\bibitem{Butterworth:2015oua}
\hrefCMSnoop {}{J.~Butterworth {et~al.}, ``{PDF4LHC} recommendations for {LHC
  Run II}'',} \textit{ J. Phys. G} \textbf{ 43} (2016) 023001,
  \href{http://dx.doi.org/10.1088/0954-3899/43/2/023001}{\doi{10.1088/0954-3899/43/2/023001}},
  \href{http://www.arXiv.org/abs/1510.03865}{\texttt{arXiv:1510.03865}}.

\bibitem{CMS:TOP-17-014}
\hrefCMSnoop {}{{CMS Collaboration}, ``Measurements of \ttbar differential
  cross sections in proton-proton collisions at $\sqrt{s}={13\TeV}$ using
  events containing two leptons'',} \textit{ JHEP} \textbf{ 02} (2019) 149,
  \href{http://dx.doi.org/10.1007/JHEP02(2019)149}{\doi{10.1007/JHEP02(2019)149}},
  \href{http://www.arXiv.org/abs/1811.06625}{\texttt{arXiv:1811.06625}}.

\bibitem{CMS:TOP-16-007}
\hrefCMSnoop {}{{CMS Collaboration}, ``Measurement of normalized differential
  \ttbar cross sections in the dilepton channel from ${\Pp\Pp}$ collisions at
  $\sqrt{s}={13\TeV}$'',} \textit{ JHEP} \textbf{ 04} (2018) 060,
  \href{http://dx.doi.org/10.1007/JHEP04(2018)060}{\doi{10.1007/JHEP04(2018)060}},
  \href{http://www.arXiv.org/abs/1708.07638}{\texttt{arXiv:1708.07638}}.

\bibitem{CMS:TOP-16-008}
\hrefCMSnoop {}{{CMS Collaboration}, ``Measurement of differential cross
  sections for top quark pair production using the lepton+jets final state in
  proton-proton collisions at {13\TeV}'',} \textit{ Phys. Rev. D} \textbf{ 95}
  (2017) 092001,
  \href{http://dx.doi.org/10.1103/PhysRevD.95.092001}{\doi{10.1103/PhysRevD.95.092001}},
  \href{http://www.arXiv.org/abs/1610.04191}{\texttt{arXiv:1610.04191}}.

\bibitem{Czakon:2017wor}
M.~Czakon\hrefCMSnoop {}{ {et~al.}, ``Top-pair production at the {LHC} through
  {NNLO} {QCD} and {NLO} {EW}'',} \textit{ JHEP} \textbf{ 10} (2017) 186,
  \href{http://dx.doi.org/10.1007/JHEP10(2017)186}{\doi{10.1007/JHEP10(2017)186}},
  \href{http://www.arXiv.org/abs/1705.04105}{\texttt{arXiv:1705.04105}}.

\bibitem{CMS:SMP-14-020}
\hrefCMSnoop {}{{CMS Collaboration}, ``Measurement of the production cross
  section of a {\PW} boson in association with two {\PQb} jets in ${\Pp\Pp}$
  collisions at $\sqrt{s}={8\TeV}$'',} \textit{ Eur. Phys. J. C} \textbf{ 77}
  (2017) 92,
  \href{http://dx.doi.org/10.1140/epjc/s10052-016-4573-z}{\doi{10.1140/epjc/s10052-016-4573-z}},
  \href{http://www.arXiv.org/abs/1608.07561}{\texttt{arXiv:1608.07561}}.

\bibitem{CMS:TOP-17-011}
\hrefCMSnoop {}{{CMS Collaboration}, ``Measurement of the single top quark and
  antiquark production cross sections in the $t$ channel and their ratio in
  proton-proton collisions at $\sqrt{s}={13\TeV}$'',} \textit{ Phys. Lett. B}
  \textbf{ 800} (2019) 135042,
  \href{http://dx.doi.org/10.1016/j.physletb.2019.135042}{\doi{10.1016/j.physletb.2019.135042}},
  \href{http://www.arXiv.org/abs/1812.10514}{\texttt{arXiv:1812.10514}}.

\bibitem{CMS:SMP-18-004}
\hrefCMSnoop {}{{CMS Collaboration}, ``${\PWp\PWm}$ boson pair production in
  proton-proton collisions at $\sqrt{s}={13\TeV}$'',} \textit{ Phys. Rev. D}
  \textbf{ 102} (2020) 092001,
  \href{http://dx.doi.org/10.1103/PhysRevD.102.092001}{\doi{10.1103/PhysRevD.102.092001}},
  \href{http://www.arXiv.org/abs/2009.00119}{\texttt{arXiv:2009.00119}}.

\bibitem{CMS:TOP-20-002}
\hrefCMSnoop {}{{CMS Collaboration}, ``Observation of ${\PQt\PW}$ production in
  the single-lepton channel in ${\Pp\Pp}$ collisions at $\sqrt{s}={13\TeV}$'',}
  \textit{ JHEP} \textbf{ 11} (2021) 111,
  \href{http://dx.doi.org/10.1007/JHEP11(2021)111}{\doi{10.1007/JHEP11(2021)111}},
  \href{http://www.arXiv.org/abs/2109.01706}{\texttt{arXiv:2109.01706}}.

\bibitem{CMS:SMP-20-014}
\hrefCMSnoop {}{{CMS Collaboration}, ``Measurement of the inclusive and
  differential ${\PW\PZ}$ production cross sections, polarization angles, and
  triple gauge couplings in ${\Pp\Pp}$ collisions at $\sqrt{s}={13\TeV}$'',}
  \textit{ JHEP} \textbf{ 07} (2022) 032,
  \href{http://dx.doi.org/10.1007/JHEP07(2022)032}{\doi{10.1007/JHEP07(2022)032}},
  \href{http://www.arXiv.org/abs/2110.11231}{\texttt{arXiv:2110.11231}}.

\bibitem{CMS:SMP-20-015}
\hrefCMSnoop {}{{CMS Collaboration}, ``Measurement of the production cross
  section for ${\PZ+\PQb}$ jets in proton-proton collisions at
  $\sqrt{s}={13\TeV}$'',} \textit{ Phys. Rev. D} \textbf{ 105} (2022) 092014,
  \href{http://dx.doi.org/10.1103/PhysRevD.105.092014}{\doi{10.1103/PhysRevD.105.092014}},
  \href{http://www.arXiv.org/abs/2112.09659}{\texttt{arXiv:2112.09659}}.

\bibitem{CMS:TOP-21-010}
\hrefCMSnoop {}{{CMS Collaboration}, ``Measurement of inclusive and
  differential cross sections for single top quark production in association
  with a {\PW} boson in proton-proton collisions at $\sqrt{s}={13\TeV}$'',}
  \textit{ JHEP} \textbf{ 07} (2023) 046,
  \href{http://dx.doi.org/10.1007/JHEP07(2023)046}{\doi{10.1007/JHEP07(2023)046}},
  \href{http://www.arXiv.org/abs/2208.00924}{\texttt{arXiv:2208.00924}}.

\bibitem{CMS:TOP-15-003}
\hrefCMSnoop {}{{CMS Collaboration}, ``Measurement of the top quark pair
  production cross section in proton-proton collisions at
  $\sqrt{s}={13\TeV}$'',} \textit{ Phys. Rev. Lett.} \textbf{ 116} (2016)
  052002,
  \href{http://dx.doi.org/10.1103/PhysRevLett.116.052002}{\doi{10.1103/PhysRevLett.116.052002}},
  \href{http://www.arXiv.org/abs/1510.05302}{\texttt{arXiv:1510.05302}}.

\bibitem{CMS:TOP-16-005}
\hrefCMSnoop {}{{CMS Collaboration}, ``Measurement of the \ttbar production
  cross section using events in the ${\Pe\PGm}$ final state in ${\Pp\Pp}$
  collisions at $\sqrt{s}={13\TeV}$'',} \textit{ Eur. Phys. J. C} \textbf{ 77}
  (2017) 172,
  \href{http://dx.doi.org/10.1140/epjc/s10052-017-4718-8}{\doi{10.1140/epjc/s10052-017-4718-8}},
  \href{http://www.arXiv.org/abs/1611.04040}{\texttt{arXiv:1611.04040}}.

\bibitem{Barlow:1993dm}
\hrefCMSnoop {}{R.~Barlow and C.~Beeston, ``Fitting using finite {Monte Carlo}
  samples'',} \textit{ Comput. Phys. Commun.} \textbf{ 77} (1993) 219,
  \href{http://dx.doi.org/10.1016/0010-4655(93)90005-W}{\doi{10.1016/0010-4655(93)90005-W}}.

\bibitem{CMS:HIG-14-009}
\hrefCMSnoop {}{{CMS Collaboration}, ``Precise determination of the mass of the
  {Higgs} boson and tests of compatibility of its couplings with the standard
  model predictions using proton collisions at 7 and {8\TeV}'',} \textit{ Eur.
  Phys. J. C} \textbf{ 75} (2015) 212,
  \href{http://dx.doi.org/10.1140/epjc/s10052-015-3351-7}{\doi{10.1140/epjc/s10052-015-3351-7}},
  \href{http://www.arXiv.org/abs/1412.8662}{\texttt{arXiv:1412.8662}}.

\end{thebibliography}\endgroup
\cleardoublepage \appendix\section{The CMS Collaboration \label{app:collab}}\begin{sloppypar}\hyphenpenalty=5000\widowpenalty=500\clubpenalty=5000
\cmsinstitute{Yerevan Physics Institute, Yerevan, Armenia}
{\tolerance=6000
A.~Tumasyan\cmsAuthorMark{1}\cmsorcid{0009-0000-0684-6742}
\par}
\cmsinstitute{Institut f\"{u}r Hochenergiephysik, Vienna, Austria}
{\tolerance=6000
W.~Adam\cmsorcid{0000-0001-9099-4341}, J.W.~Andrejkovic, T.~Bergauer\cmsorcid{0000-0002-5786-0293}, S.~Chatterjee\cmsorcid{0000-0003-2660-0349}, K.~Damanakis\cmsorcid{0000-0001-5389-2872}, M.~Dragicevic\cmsorcid{0000-0003-1967-6783}, A.~Escalante~Del~Valle\cmsorcid{0000-0002-9702-6359}, P.S.~Hussain\cmsorcid{0000-0002-4825-5278}, M.~Jeitler\cmsAuthorMark{2}\cmsorcid{0000-0002-5141-9560}, N.~Krammer\cmsorcid{0000-0002-0548-0985}, L.~Lechner\cmsorcid{0000-0002-3065-1141}, D.~Liko\cmsorcid{0000-0002-3380-473X}, I.~Mikulec\cmsorcid{0000-0003-0385-2746}, J.~Schieck\cmsAuthorMark{2}\cmsorcid{0000-0002-1058-8093}, R.~Sch\"{o}fbeck\cmsorcid{0000-0002-2332-8784}, D.~Schwarz\cmsorcid{0000-0002-3821-7331}, M.~Sonawane\cmsorcid{0000-0003-0510-7010}, S.~Templ\cmsorcid{0000-0003-3137-5692}, W.~Waltenberger\cmsorcid{0000-0002-6215-7228}, C.-E.~Wulz\cmsAuthorMark{2}\cmsorcid{0000-0001-9226-5812}
% ...
\par}
\vskip\cmsinstskip
\dag:~Deceased\\
$^{1}$Also at Yerevan State University, Yerevan, Armenia\\
$^{2}$Also at TU Wien, Vienna, Austria\\
% ...
\end{sloppypar}
%%% END EDITABLE REGION %%%
% skeleton_end
\end{document}